\documentclass[twocolumn]{aastex631}

\usepackage{derivative}
\usepackage{amsmath}
\usepackage{ulem}
\usepackage{tabularx}
\usepackage{longtable}
\usepackage{threeparttable}
\usepackage{enumerate}
\usepackage{natbib}
\usepackage{multirow}
\usepackage{appendix}

\usepackage[utf8]{inputenc}
\usepackage[T1]{fontenc}
\usepackage{courier}  % Courier (monospaced)

\accepted{March 1, 2025}
\submitjournal{ApJ}

\shorttitle{Formation of Cold Jupiters}
\shortauthors{Guo et al.}
\graphicspath{{./}{figures/}}
\begin{document}

\title{A Population Synthesis Study on the Formation of Cold Jupiters from Truncated Planetesimal Disks}

\correspondingauthor{Kangrou Guo}
\email{carol.guo@sjtu.edu.cn}

\author[0000-0001-6870-3114]{Kangrou Guo}
\affiliation{Tsung-Dao Lee Institute, Shanghai Jiao Tong University, 1 Lisuo Road, Shanghai 201210, China}

\author[0000-0002-8300-7990]{Masahiro Ogihara}
\affiliation{Tsung-Dao Lee Institute, Shanghai Jiao Tong University, 1 Lisuo Road, Shanghai 201210, China}
\affiliation{School of Physics and Astronomy, Shanghai Jiao Tong University, 800 Dongchuan Road, Shanghai 200240, China}

\author[0000-0001-9564-6186]{Shigeru Ida}
\affiliation{Earth-Life Science Institute, Institute of Science Tokyo, Meguro, Tokyo 152-8550, Japan
}

\author[0000-0003-4676-0251]{Yasunori Hori}
\affiliation{Astrobiology Center, Osawa 2-21-1, Mitaka, Tokyo 181-8588, Japan}
\affiliation{National Astronomical Observatory of Japan, Osawa 2-21-1, Mitaka, Tokyo 181-8588, Japan
}

\author[0000-0003-1535-5587]{Kaiming Cui}
\affiliation{Department of Physics, University of Warwick, Coventry CV4 7AL, UK}
\affiliation{Centre for Exoplanets and Habitability, University of Warwick, Gibbet Hill Road, Coventry CV4 7AL, UK}

\author[0000-0001-6039-0555]{Fabo Feng}
\affiliation{Tsung-Dao Lee Institute, Shanghai Jiao Tong University, 1 Lisuo Road, Shanghai 201210, China}
\affiliation{School of Physics and Astronomy, Shanghai Jiao Tong University, 800 Dongchuan Road, Shanghai 200240, China}

\begin{abstract}

The occurrence rate of giant planets increases with orbital period and turns over at a location that roughly corresponds to the snow line of solar-type stars. Further, the density distribution of cold Jupiters (CJs) on the semi-major axis - mass diagram shows a relatively steep inner boundary, shaping the desert of warm Jupiters. The eccentricities of CJs show a broad distribution with a decreasing number density towards the larger end. Previous planet formation models fail to reproduce all these features at the same time. We use a planet population synthesis (PPS) model with truncated initial planetesimal distribution and compare the mass and orbital distribution of the simulated planets with the observation. We show that the occurrence of CJs with respect to the orbital period, the slope of the inner boundary of CJs on the semi-major axis - mass diagram, and the eccentricity distribution of CJs agree reasonably well with observation, if CJs form from truncated planetesimal disks of 10 au or wider with suppressed migration. While PPS simulations generally overestimate the fraction of giants with eccentricity below 0.2, $N$-body simulations produce a more consistent eccentricity distribution with observation. While the fraction of high-eccentricity planets can be increased by widening the planetesimal disk or reducing the migration speed, a deficit of giants with eccentricity between 0.2—0.4 exists regardless of the choices of parameters. Our results indicate that CJs are more likely born in truncated disks near the snow line than in classical uniform disks.

\end{abstract}

% \keywords{Classical Novae (251) --- Ultraviolet astronomy(1736) --- History of astronomy(1868) --- Interdisciplinary astronomy(804)}

\section{Introduction} \label{sec:intro}

Cold gas giant planets, commonly referred to as cold Jupiters (CJs), are typically defined as planets with masses greater than 0.3 Jupiter masses ($M_{\rm{J}}$) and orbital distances beyond 1 au \citep{Zhu_2018}. Detecting CJs is more challenging compared to closer-in planets due to their long orbital periods. %, which require extended observational baselines in radial velocity (RV) surveys. 
However, advancements in observational instruments and techniques in recent years have significantly increased the number of observed CJs. The 8-year radial velocity (RV) data from HARPS, along with additional data from CORALIE, have expanded the detectable orbital distances of planets to nearly 10 au \citep{Mayor_2011}. Additionally, the California Legacy Survey \citep[e.g.,][]{Fulton_2021} has reported giant planets with orbits extending beyond 10 au. Using combined analyses of RV, astrometry, and direct imaging data, \citet{Feng_2022} identified 167 cold giant planets along with other types of substellar companions, further extending the sample of the observed distant giant planets.
This growing body of CJ observational data offers valuable insights for constraining the theories of their formation and evolution.

The distribution of exoplanets in the semi-major axis–mass $(a,m)$ diagram offers essential insights into their formation and evolutionary history. Examining the distribution of giant planets detected via RV in this diagram, we observe an inner boundary with an approximate slope of 3 (see Section \ref{sec:inner_boundary_obs} for details). This boundary shapes the region where the occurrence of warm Jupiters (WJs, here we refer to giant planets with $0.1< a< 1$ au) is low. Since this boundary conveys two-dimensional information—both mass and orbital distance—about the distribution of giant planets, it is a key factor in understanding their growth and migration histories. 

The occurrence of giant planets with respect to the orbital period is another important observational clue to understand the formation of CJs. 
It has been shown that the occurrence of giant planets within the mass range of 0.1-20 $M_{\rm{J}}$ increases with the orbital period and turns over at a location that roughly corresponds to the snow line of sun-like stars \citep{Fernandes_2019} [hereafter F19].
Such a pile-up of CJs near the snow line indicates that they either formed with particularly high efficiency near this location, or stalled their migration there after forming on wider orbits.

Beyond orbital distances from the central star, the eccentricities of giant planets are key indicators of their dynamical history. Using RV data, \citet{Kane_2024} reports a median eccentricity of 0.23 for cold giants. Specifically, \citet{Rosenthal_2024} finds that single giant planets in the CLS sample typically have near-circular orbits, though their eccentricity distribution includes a long tail extending to high values (up to $e=0.77$). In contrast, multiple giant planets generally exhibit moderate eccentricities, with their distribution extending to a lower maximum ($e=0.47$ with 90th percentile). By filtering RV-detected giant planets with semi-major axes larger than 0.1 au, we observe a broad eccentricity distribution, with number density decreasing toward higher values. Formation models of CJs must therefore produce eccentricity distributions consistent with these observations.

Planet population synthesis (PPS) simulations are a powerful tool for revealing the correlations between initial conditions (such as star and disk properties) and the final outcomes of planet formation. One of the pioneering works is the Ida \& Lin model (hereafter the IL model) by \citet{Ida_2004a, Ida_2004b}. Over time, the IL model has been updated to include type I migration \citep{Ida_2008a}, giant impacts between planetary embryos \citep{Ida_2010}, scattering of giant planets \citep{Ida_2013}, and type II migration with a two-$\alpha$ disk model \citep{Ida_2018}.
However, the latest version of the IL model has limitations in reproducing the observed distribution of giant planets—such as the inner boundary on the $(a,m)$ diagram, the occurrence rate across orbital periods, and eccentricity. For instance, F19 found that simulations using the IL model \citep{Ida_2018} produce a much flatter slope in the occurrence distribution of giant planets with respect to orbital period compared to observations, meaning the model over-predicts the number of giant planets with periods shorter than $\sim 1000$ days. Furthermore, when comparing the giant planet distribution on the $(a,m)$ diagram produced by the IL model with RV data, no inner boundary consistent with observations was found.
Another advanced PPS model is the Generation III Bern (NGPPS) model \citep[e.g.,][]{Emsenhuber_2021_I, Emsenhuber_2021_II}. In addition to simulating planet formation stages, the NGPPS model incorporates the thermodynamic evolution of internal planetary structures and tidal migration in the post-formation stages. While the 20-core Bern model better reproduces the giant planet occurrence rate with respect to orbital period than the IL model (F19), it does not generate an inner boundary of giant planets on the $(a,m)$ diagram that matches RV observations.
Gas giant planet formation via pebble accretion has also been extensively studied and has often been raised as a potential solution to retaining distant giant planets owing to the high accretion efficiency of planetary cores \citep[e.g.,][]{Ormel_2010, Lambrechts_2012, Morbidelli_2012, Bitsch_2015, Liu_2019b}.
Cold giant planets can grow from distant embryos ($a \gtrsim 15$ au, \citealt{Chambers_2018, Ndugu_2018, Bitsch_2019, Johansen_2019}) where growth by planetesimal collision is less efficient. 
%The pebble accretion efficiency can also be limited when pebble erosion is more efficient than dust sticking in the envelope of the growing planet \citep{Ali-Dib_2020}.
However, regardless of pebble accretion or planetesimal accretion, classic planet formation models generally produce too many WJs, fuzzing up the inner boundary of CJ on the $(a,m)$ diagram. 
Regarding eccentricity distributions, many planet formation models tend to overestimate the fraction of stable systems, leading to an excess of low-eccentricity ($e\lesssim 0.2$) planets compared to what is observed \citep{Ida_2013, Matsumura_2021}.
\citet{Bitsch_2019} identifies two typical outcomes of orbital instability in systems of multiple giant planets: systems either undergo dramatic instabilities, leaving behind only two highly eccentric giants, or they experience minor instability with several planets remaining on nearly circular orbits. A systematic investigation of the orbital and mass distribution of giant planets is therefore needed.

Classical planet formation models often assume a power-law profile for gas and solid distributions, as described in the minimum mass solar nebula model (MMSN; \citealt{Hayashi_1981}). 
However, several theories have predicted that planetesimals can form in discrete locations instead of smoothly distributed in the disk.
For example, the pile-up of pebbles at the pressure maxima in the disk can be possible sites of planet formation and likely cause the inside-out planet formation from pebble rings \citep{Chatterjee_2014}.
Considering dust growth and radial drift in a standard viscous disk model, \citet{Drazkowska_2016} shows that the global redistribution of solids can lead to a pile-up of pebbles in the inner disk, possibly resulting in planetesimal formation in a narrow ring near 1 au. 
In addition, the snow line is often considered a critical location for the accumulation of dust or pebbles and a natural site for planetesimal formation \citep{Ros_2013, Ida_2016, Schoonenberg_2017, Drazkowska_2017, Hyodo_2019, Guilera_2020, Charnoz_2021, Morbidelli_2022}. 
Several models have successfully reproduced terrestrial planets formed from discrete planetesimal rings, demonstrating consistency with observations of both the Solar System \citep{Hansen_2009, Ogihara_2018, Ueda_2021, Izidoro_2022, Woo_2023, Woo_2024} and exoplanetary systems \citep{Batygin_2023, Ogihara_2024}. \citet{Izidoro_2022} demonstrated that the growth of Jupiter and Saturn from a wide planetesimal ring, located between approximately 3--4 au and 10--20 au, which formed due to dust accumulation at the water snow line, produces a system architecture consistent with that of the Solar System. However, the formation of gas giant planets from planetesimal rings, and its comparison with observed extrasolar giant planets, remains a relatively under-explored topic. To address this gap, we investigate the formation of giant planets from planetesimal rings produced near the water snow line (due to the relatively large width of the rings, we use the term "truncated planetesimal disks" throughout this paper). 

In this study, we particularly aim to reproduce the density distribution on the $(a,m)$ diagram, the occurrence rate with respect to orbital period, and the eccentricity distribution of CJs using a truncated disk model. 
This paper is organized as follows: Section \ref{sec:inner_boundary_obs} explains how the inner boundary of giant planets on the $(a,m)$ diagram is determined by the density distribution. Section \ref{sec:method} outlines the planet formation model used, along with the details of our model setup and simulations. In Section \ref{sec:a-m_distribution}, we present our PPS simulation results, focusing on the $(a,m)$ distribution and the occurrence rate as a function of orbital period. Section \ref{sec:eccentricity} shows the eccentricity distribution of giant planets produced by our models. In Section \ref{sec:discussion} we discuss the results and the implications on planet formation. Finally, we summarize our findings in Section \ref{sec:conclusion}.

\section{Inner boundary in semi-major axis - mass diagram}
\label{sec:inner_boundary_obs}

\begin{figure*}
    \centering
    \includegraphics[width=0.95\textwidth]{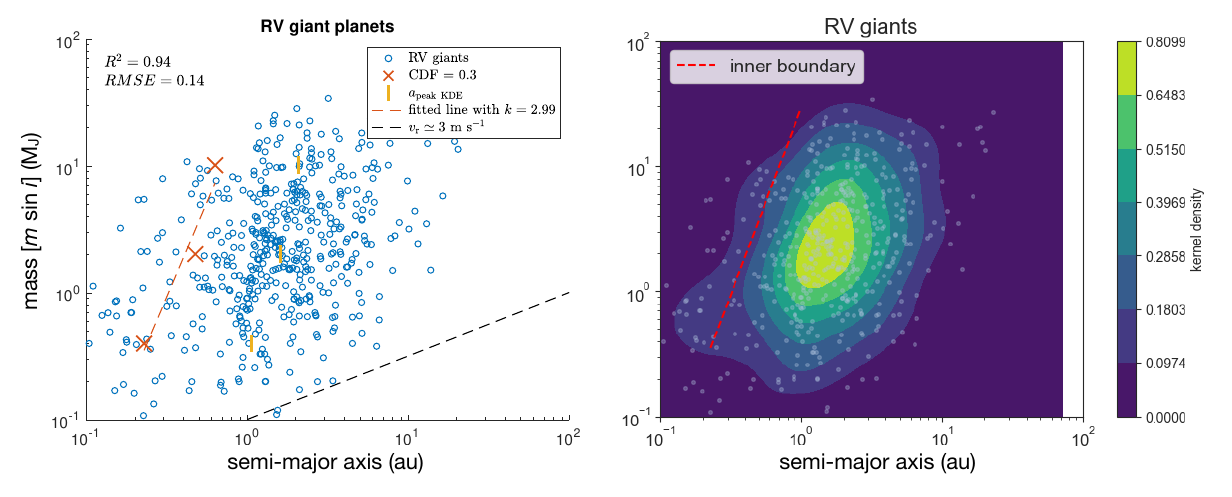}
    \caption{\textit{Left:} scatter plot of giant planets in the RV sample on the $(a,m)$ diagram. The yellow vertical bars mark the locations $a_{\rm{peak,KDE}}$ that correspond to the maximum kernel density value in each mass bin. The red crosses represent the locations of the inner boundary where the cumulative kernel density within 0.1 au and $a_{\rm{peak,KDE}}$ is 0.3. The inner boundary is then fitted by the red dashed line. The black dashed line shows the RV detection limit where the radial velocity $v_{\rm{r}} \simeq 3~\rm{m}~s^{-1}$ for reference. \textit{Right:} two-dimensional KDE plot of RV giant planets on the $(a,m)$ diagram. The same inner boundary is plotted with the red dashed line as in the left panel.}
    \label{fig:RV_obs}
\end{figure*}

The inner boundary of CJ distribution on the $(a,m)$ diagram provides rich information of the growth and migration history of giant planets. In this section, we describe how we identify and quantify such an inner boundary using a composite RV sample acquired from the NASA Exoplanet Archive \citep{PSCompPars}. 
Because transit observations are heavily biased towards close-in planets, we use the data of planets observed by RV method only. 
We download the data of planets that are (1) detected by RV method (2) in single star systems (2) with known host mass and metallicity (3) with known mass ($m\sin{i}$), semi-major axis, eccentricity, and orbital period. 
Then we select the planets with host mass smaller than 3 solar mass ($M_* < 3~M_\odot$), due to the high observational uncertainties around massive stars. 
For planets with multiple entries listed, we keep the data of the latest detection only.
To focus on cold giant planets, we set a mass range of $m~\sin{i}>0.1~M_{\rm{J}}$ and a semi-major axis range of $a>0.1$ au (although CJ typically refers to giant planets beyond 1 au, we loosen the constraints on the orbital distance when filtering the planets in order to see the boundary that separates CJ from WJ). 
This gives us 483 giant planets around 402 stars. The mass and metallicity distribution of the 402 stars are shown in Figure \ref{fig:RV_mass_metallicity}.
%\citet{Mayor_2011} (hereafter referred to as the Mayor+11 sample)}, which includes 822 stars and 155 planets combining the HARPS and CORALIE surveys.

The left panel in Figure \ref{fig:RV_obs} shows the scatter plot of the planets in the RV sample on the semi-major axis - mass $(a,m)$ diagram. The inner boundary, which is highlighted by the red dashed line, is determined by the density distribution.
%\textbf{To account for the detection efficiency, we apply the inverse detection completeness in \citet{Mayor_2011} as the weights when evaluating the density distribution of the RV planets (see Appendix \ref{appendix:RV_reliability} for a discussion on the robustness of this treatment).
%The detection efficiency in \citet{Mayor_2011} is available online and included in the \texttt{epos} package \citep{Mulders_2018, Fernandes_2019}.}
Three mass bins are created in logarithmic scale within the range of [0.1, 20] $M_{\rm{J}}$. % This is to avoid the large uncertainties of the density estimation due to the small number of data points outside this mass range. 
In order to eliminate the influence of the lower detection efficiency for the planets with larger orbital distance on the density calculation, for each mass bin we first calculate the semi-major axis $a_{\rm{peak,KDE}}$ where the kernel density estimates (KDE) for the probability distribution reaches the maximum (marked by the yellow vertical bars). 
The KDE is calculated using a Gaussian function as the kernel smoother and a bandwidth of 0.4. %The detection completeness \citep{Mayor_2011} is applied as the weights.
Then we calculate the cumulative distribution $K_{\rm{cumulative}}$ of planets that fall within 0.1 au and $a_{\rm{peak,KDE}}$. 
We set the location where $K_{\rm{cumulative}} = 0.3$ as the inner boundary $a_{\rm{boundary}}$ (marked by the red crosses).
Finally, we fit the values of $a_{\rm{boundary}}$ in each bin with a 1st order polynomial (the red dashed line). The slope of the fitted line is calculated as the slope of the inner boundary $k$. 
We check the quality of the fitting by calculating the coefficient of determination $R^2$ and the root mean square error $RMSE$. The two quantities are given by $R^2 = 1-\Sigma(y_i-\hat{y}_i)^2/\Sigma(y_i-\bar{y})^2$ and $RMSE = \sqrt{\Sigma(y_i-\hat{y}_i)^2/n}$, where $y_i$ are the original data points, $\hat{y}_i$ are the fitted values, $\bar{y}$ is the mean of the original data points, and $n$ is the number of data points. We adjust the width and number of the mass bin so that the $R^2$ value is the largest (this is equivalent to ensuring the lowest $RMSE$ value).
In recent years, RV surveys can reach a precision of $v_{\rm{r}}$ on the order of a few m s$^{-1}$ \citep[e.g.,][]{Fulton_2021}, where $v_{\rm{r}}$ is the radial velocity variation of the star induced by the planet. Therefore, we draw a line that corresponds to $v_{\rm{r}} \simeq 3~\rm{m}~s^{-1}$ for reference. Planets below this line are very difficult to be detected by RV due to long orbital periods and small masses.

The right panel in Figure \ref{fig:RV_obs} shows the two-dimensional KDE plot of the RV sample. 
The same bandwidth as above is applied in the density calculation.
%We apply the observation completeness in \citet{Mayor_2011} to our selected sample, and weight the KDE by the observation completeness. 
The inner boundary calculated as described above is over-plotted as the red dashed line. The fitted inner boundary shows a good agreement with the contours in the 2D KDE plot. 
Therefore, we find that the inner boundary of CJ distribution on the $(a,m)$ diagram has a slope of $k\simeq 3$ (see Appendix \ref{appendix:RV_reliability} for a comparison with inner boundaries derived from more conservative samples with controlled biases). 

We also used the occurrence rate as a function of planet mass and orbital period given in Equation (3) in F19 to calculate the inner boundary slope. Using the symmetric \texttt{epos} parameters $p_1=0.65$ and $m_1=-0.45$ given in their Table 1, we obtain an inner boundary, along which the occurrence rate is a constant, with a slope of 2.17. This slope is qualitatively consistent with our calculation from the 2D density distribution.

\section{Method} \label{sec:method}

We build an updated planet population synthesis (PPS) model (hereafter the new model) based on the IL model \citep[e.g.,][]{Ida_2013, Ida_2018}.

%The workflow of our numerical simulations is shown in Figure \ref{fig:workflow}.
The initial conditions including the stellar and disk properties are given as input for the PPS simulations. 
For the purpose of demonstrating the distribution of giant planets on the $(a,m)$ diagram and the occurrence rate along the orbital period (as in Section \ref{sec:a-m_distribution}), we use the results from PPS simulations only for the sake of a large sample size. For investigating the eccentricity distribution of giant planets (as in Section \ref{sec:eccentricity}), we use results from PPS simulations integrated with $N$-body simulations for more accurate eccentricity estimates.
The details are presented as follows.

\subsection{Planet formation model} \label{subsec:PPS_model}

Here we summarize the framework of the IL model, and highlight the modifications made for the new model. 
For more details of the IL model, we refer to \cite{Ida_2013} and \cite{Ida_2018}.

\subsubsection{Disk model}

For the gas component of the disk, we follow \citet{Ida_2013} and adopt the $r$-dependence of the steady accretion disk with constant $\alpha$ viscosity as
\begin{equation}
    \Sigma_{\rm{g}} = \Sigma_1 f_{\rm{g}}\left(\frac{r}{1~\rm{au}}\right)^{-1},
    \label{eq:Sigma_g}
\end{equation}
where $r$ is the disk radius, $\Sigma_1 = 750~\rm{g}~{cm}^{-2}$ is the gas surface density at 1 au and $f_{\rm{g}}$ is a scaling factor to adjust the mass of the gas disk.

For the solid component, we consider a planetesimal ring with a given width $\Delta a$. 
During the Class II stage of disk evolution, the snow line gradually migrates inward \citep[e.g.,][]{Garaud_2007, Oka_2011, Drazkowska_2018, Lichtenberg_2021}, potentially extending the planetesimal ring over a certain width and creating a truncated planetesimal disk at the snow line.
For simplicity, we use a power-law distribution as in the MMSN model to describe the solid surface density within the truncated planetesimal disk, and vary the disk width as a free parameter.
The inner edge of the disk is located at the snow line $a_{\rm{ice}}$, and the outer edge is located at $a_{\rm{ice}}+\Delta a$.
The location of the snow line in an optically thin disk is given by $a_{\rm{ice}} = 2.7 (L_*/L_\odot)^{1/2}$ au \citep{Hayashi_1981}, where $L$ is the stellar luminosity. We adopt the relation given in \cite{Ida_2005} as 
\begin{equation}
    a_{\rm{ice}} = 2.7 \left(\frac{M_*}{M_\odot}\right)^2~\rm{au}.
    \label{eq:a_snow}
\end{equation}
Therefore, the solid surface density of the truncated planetesimal disk is 
\begin{equation}
    \Sigma_{\rm{d}} = f_{\rm{d}} \left(\frac{r}{1~\rm{au}}\right)^{-1.5}, \quad a_{\rm{ice}} \le r < a_{\rm{ice}}+\Delta a,
    \label{eq:Sigma_d}
\end{equation}
where $f_{\rm{d}}$ is a scaling factor to control the total mass in the disk $m_{\rm{tot}} = \int_{a_{\rm{ice}}}^{a_{\rm{ice}}+\Delta a} 2\pi r \Sigma_{\rm{d}} dr$.

\begin{figure}
    \centering
    \includegraphics[width=0.47\textwidth]{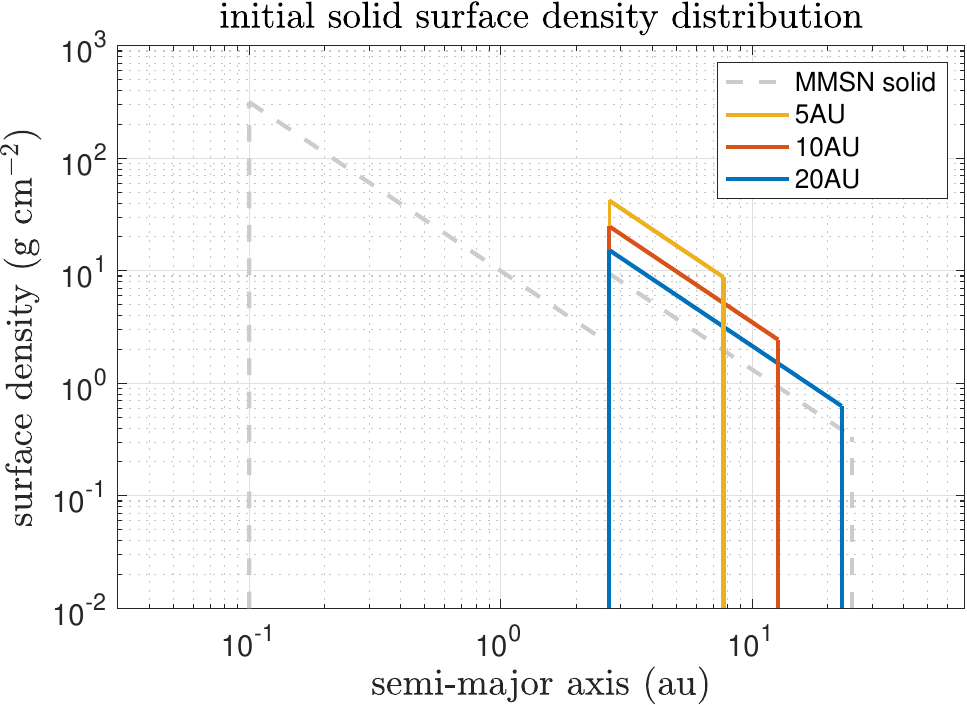}
    \caption{The initial solid surface density distribution for $M_*=1~M_\odot$ case. The grey dashed line shows the solid surface density given by the MMSN model. The yellow, red, and blue solid lines represent the solid surface density distribution in the truncated disk model, each with a different disk width while the total solid mass is fixed as 100 $M_\oplus$.}
    \label{fig:surface_density}
\end{figure}

Figure \ref{fig:surface_density} shows the initial solid surface density distribution in different models for comparison (when $M_*=1~M_\odot$). The grey dashed line shows the solid surface density distribution as in the MMSN model. The yellow, red, and blue solid lines show the solid surface density profile in the 5-au, 10-au, and 20-au truncated disk models (Models 1--3), respectively. The total masses within the three truncated disks are fixed as $100~M_\oplus$. 

The self-similar solution for disk evolution is adopted \citep{Lyndel-Bell_1974}
\begin{equation}
    \Sigma(r,t) = \Sigma_0\left(\frac{r}{r_0}\right)^{-1} \tilde{t}^{-3/2}\exp{\left(-\frac{r}{\tilde{t}r_0}\right)},
    \label{eq:self_similar}
\end{equation}
where $\tilde{t} = (t/t_{\rm{dep}})+1$, $t_{\rm{dep}}$ is the disk depletion timescale, $r_0$ is the initial disk radius, and $\Sigma_0$ is the initial surface density.

The disk accretion rate is 
\begin{equation}
    \dot{M}_{\rm{g}} \sim 3 \pi \Sigma \nu_{\rm{acc}} \sim 3\pi \Sigma \alpha_{\rm{acc}}\left(\frac{h}{r}\right)^2r^2\Omega, \quad r \ll r_0,
    \label{eq:m_dot_disk}
\end{equation}
where $\nu_{\rm{acc}} = \alpha_{\rm{acc}}(h/r)^2r^2\Omega$ is the effective kinetic viscosity, $h$ is the scale height of the gas disk, $\Omega$ is the Keplerian frequency, and $\alpha_{\rm{acc}}$ is the alpha parameter for disk accretion.
Here we adopt the value $\alpha_{\rm{acc}}\sim10^{-3}$.

One update we add to the IL model is to adopt two disk depletion timescales to mimic the rapid disk clearing by disk winds and photo-evaporation at late stages \citep[e.g.,][]{Izidoro_2015, Ogihara_2020}. Following \cite{Hori_2020}, we adopt a disk dispersal timescale of $\tau_{\rm{disk}} \simeq 2.5$ Myr, and introduce a late-stage rapid disk clearing timescale of $\tau_{\rm{rapid}} \simeq 10$ kyr.
Therefore when calculating the time evolution of the gas surface density in Equation \ref{eq:self_similar}, we adopt $t_{\rm{dep}} = \tau_{\rm{disk}}~(t \leq \tau_{\rm{disk}})$ and $t_{\rm{dep}} = \tau_{\rm{rapid}}~(t >\tau_{\rm{disk}})$.
%\todo{may insert a figure here showing the two disk accretion rates?}

\subsubsection{Core accretion}
The core accretion timescale is calculated by \citep{Kokubo_Ida_2002}
\begin{align}
    \tau_{\rm{c,acc}} \simeq &1.2 \times 10^5 \left(\frac{\Sigma_{\rm{d}}}{10~{\rm{g}~cm^{-2}}}\right)^{-1} \left(\frac{a}{1~\rm{au}}\right)^{1/2} \left(\frac{M_{\rm{c}}}{M_{\oplus}}\right)^{1/3} \nonumber \\
    &\times \left(\frac{M_*}{M_{\odot}}\right)^{-1/6} \left[\left(\frac{\Sigma_{\rm{g}}}{2.4\times10^3~\rm{g}~cm^{-2}}\right)^{-1/5} \right. \nonumber \\
    &\times \left. \left(\frac{a}{1~\rm{au}}\right)^{1/20}\left(\frac{m_{\rm{p}}}{10^{18}~\rm{g}}\right)^{1/15}\right]^2~\rm{yr},
    \label{eq:t_c,acc}
\end{align}
where $M_{\rm{c}}$ is the core mass and $m_{\rm{p}}$ is the mass of the planetesimal, which we adopt as $m_{\rm{p}}=10^{18}$g in the PPS simulations.

In the ideal case where there is no radial migration of the core, the core grows until it consumes all planetesimals in its feeding zone and reaches the core isolation mass, which is given by
\begin{align}
    M_{\rm{c,iso}} \simeq & 0.16 \left(\frac{\Sigma_{\rm{d}}}{10~\rm{g}~cm^{-2}}\right)^{3/2}\left(\frac{a}{1~\rm{au}}\right)^3 \nonumber \\
    & \times \left(\frac{\Delta a_{\rm{c}}}{10r_{\rm{H}}}\right)^{3/2}\left(\frac{M_*}{M_{\odot}}\right)^{-1/2}~M_{\oplus},
    \label{eq:m_c,iso}
\end{align}
where $\Delta a_{\rm{c}}$ is the orbital separation between the cores, and $r_{\rm{H}}$ is the Hill radius $r_{\rm{H}} \equiv \left(\frac{M_{\rm{c,iso}}}{3M_*}\right)^{1/3}a$.
In the PPS simulation, the concurrent growth and migration of all cores are integrated simultaneously with the $\Sigma_{\rm{d}}$ and $\Sigma_{\rm{g}}$  distribution, which evolve through time. Multiple cores (planet seeds with a mass of $10^{24}$ g) are initially distributed within the truncated disks.
% We assume that the runaway growth phase quickly changes to the oligarchic growth phase (Kokubo & Ida 1998). Since in the oligarchic growth phase, the core's accretion timescale is dominated by their late-stage growth ($\tau_{\rm{c,acc}} \propto M_{\rm{c}}^{1/3}$), the dependence on their initial mass is negligible \citep{Ida_2008}.
In disk regions where the growth timescale is short, the initial separation between cores is set to be the full feeding-zone width of the local asymptotic isolation mass ($\Delta a_{\rm{c}} = 10 r_{\rm{H}}(M_{\rm{c,iso}}$)). When $M_{\rm{c,iso}} > 10~M_\oplus$, the separation is set to $10 r_{\rm{H}}(10M_\oplus)$ instead to avoid unreasonably large spacing. In more distant disk regions where the cores are unlikely to reach their isolation mass within the life span of their host stars, the embryos are separated by the feeding zone width of the mass evaluated for the local $\Sigma_{\rm{d}}$ after $t\sim 1$ Gyr. 
% If the effect of radial migration is taken in to account, the final core mass can be larger than $M_{\rm{c,iso}}$.
% In the extreme case where an inwardly migrating core can accrete all planetesimals inside its orbit, the core would acquire a maximum asymptotic mass of 
% \begin{equation}
%     M_{\rm{c,no}~iso} \sim \pi a^2 \Sigma_{\rm{d}} \simeq 1.2 \left(\frac{\Sigma_{\rm{d}}}{10~\rm{g}~cm^{-2}}\right)\left(\frac{a}{1~\rm{au}}\right)^2~M_{\oplus}.
%     \label{eq:m_c,noiso}
% \end{equation}

\subsubsection{Gas accretion and migration}

The critical core mass for initiating gas accretion is given by 
\begin{equation}
    M_{\rm{c,crit}} \simeq 10 \left(\frac{\dot{M}_{\rm{c}}}{10^{-6}M_{\oplus}~\rm{yr}^{-1}}\right)^{0.25}~M_{\oplus},
    \label{eq:m_crit}
\end{equation}
where $\dot{M}_{\rm{c}} = M_{\rm{c}}/\tau_{\rm{c,acc}}$ is the core accretion rate \citep{Ikoma_2000, Hori_2010}.
Here the dependence of $M_{\rm{c,crit}}$ on the opacity of the dust grain $\kappa$ is neglected for simplicity, and we assume that $\kappa = 1~\rm{cm}^2~g^{-1}$.
After reaching the critical core mass, the planet mass increase is regulated by the Kelvin-Helmholtz contraction timescale $\tau_{\rm{KH}}$ as
\begin{equation}
    \frac{dM_{\rm{p}}}{dt} \simeq \frac{M_{\rm{p}}}{\tau_{\rm{KH}}},
    \label{eq:KH_growth}
\end{equation}
where $M_{\rm{p}}$ is the total mass of the planet including its gas envelope, and the Kelvin-Helmholtz timescale is given by $\tau_{\rm{KH}} \simeq 10^8\left(\frac{M_{\rm{p}}}{M_{\oplus}}\right)^{-3.5}$ years.

During the core growth, imbalance of the inner and outer torques exerted by the gas disk on the core causes the type I migration of the core \citep{Goldreich_1980, Artymowicz_1993}. For a planet core at a radial distance $r$, we adopt the timescale of type I migration given by three-dimensional linear calculation \citep{Tanaka_2002,Kanagawa_2018}:
\begin{equation}
    \tau_{\rm{mig,I}} = \frac{1}{2f_1}\left(\frac{M_{\rm{p}}}{M_*}\right)^{-1}\left(\frac{\Sigma_{\rm{g}} r^2}{M_*}\right)^{-1}\left(\frac{h}{r}\right)^2\Omega^{-1}.
    \label{eq:tau_0}
\end{equation}
Following \citet{Ida_2008a}, we use a scaling factor $f_1$ to account for the retardation of type I migration caused by nonlinear effects. Possible retardation processes could include variations in the disk surface density and temperature gradient \citep{Masset_2006b}, MHD-driven turbulence \citep{Laughlin_2004, Nelson_2004}, self-excited vortices \citep{Li_2005}, and nonlinearities of the flow around the planet \citep{Masset_2006a}. 
% need to add more recent references
By employing the wind-driven disk model \citep{Suzuki_2010} and using the torque formula in \citet{Paardekooper_2011}, \citet{Ogihara_2015a,Ogihara_2015b} have shown that Type I migration in windy disks could be strongly reduced due to the modified surface density slope induced by the disk wind.
In addition to the interaction with the gas component of the disk, \citet{Hsieh_2020} found that the dusty dynamical corotation torques can also slow down the migration of low-mass planets in dust-rich disks.
To account for these uncertainties, we adopt $f_1$ as a free parameter in the model and examine how the results depend on the migration efficiency. %\todo{Any other missed-out references here?}

When the planet becomes massive enough to perturb the local gaseous disk, it opens a (partial) gap in the protoplanetary disk.
A gap is opened by the planet at an orbital distance of $r$ when both the thermal condition
\begin{equation}
    h < r_{\rm{H}} = \left(\frac{M_{\rm{p}}}{3M_*}\right)^{1/3}r
    \label{eq:thermal_condition}
\end{equation}
and the viscous condition
\begin{equation}
    \frac{M_{\rm{p}}}{M_*} > \frac{40\nu_{\rm{vis}}}{r^2\Omega}
    \label{eq:viscous_condition}
\end{equation}
are satisfied \citep{Lin_1993,Crida_2006}. 
Here $\nu_{\rm{vis}} = \alpha_{\rm{vis}}(h/r)^2r^2\Omega$ is the turbulent viscosity, since gap opening is controlled local disk turbulence, which is expected to be very weak according to non-ideal MHD simulations \citep[e.g.,][]{Bai_2017}.
Following \citet{Ida_2018}, we adopt $\alpha_{\rm{vis}} \sim 0.1 \alpha_{\rm{acc}}$.

After gap opening, planet migration switches from Type I to Type II, and the gas accretion rate is limited by the disk supply 
\begin{equation}
    \frac{dM_{\rm{p}}}{dt} \simeq \min{\left[\frac{dM_{\rm{p,KH}}}{dt},\dot{M}_{\rm{g}} \exp{\left(-\frac{M_{\rm{p}}}{M_{\rm{p,th}}}\right)f_{\rm{lc}}}\right]},
\end{equation}
where $M_{\rm{p,th}}$ is the planet mass when $r_{\rm{H}}\simeq 2h$ ($M_{\rm{p,th}} \sim 120(r/1{\rm{au}})^{3/4}M_\oplus$),
$f_{\rm{lc}}$ is the reduction factor for the local gas surface density due to gap opening \citep{Tanaka_2020}
\begin{equation}
    f_{\rm{lc}} = \frac{D'/3\pi \nu_{\rm{vis}}}{1+D'/3\pi\nu_{\rm{vis}}},
    \label{eq:f_lc}
\end{equation}
where $D'$ is given by
\begin{equation}
    D' = \frac{D}{1+0.04K},
\end{equation}
and \citep{Kanagawa_2018}
\begin{eqnarray}
    D &=& 0.29 \left(\frac{M_{\rm{p}}}{M_*}\right)^{4/3}\left(\frac{h}{r}\right)^{-2}r^2\Omega, \\
    K &=& \left(\frac{M_{\rm{p}}}{M_*}\right)^2\left(\frac{h}{r}\right)^{-5} \alpha_{\rm{vis}}^{-1}.
\end{eqnarray}

The timescale of type II migration is then given by
\begin{equation}
    \tau_{\rm{mig,II}} = -\frac{1+0.04K}{\gamma_L+\gamma_C\exp{(-K/K_{\rm{t}})}} \tau_{\rm{mig,I}}.
    \label{eq:t_mig,II}
\end{equation}
%where $f_1$ is the retardation factor for type I migration, $\tau_0$ is the timescale of type I migration (without a gap)
Following \citet{Kanagawa_2018}, we adopt $\gamma_L = -3.1119$, $\gamma_C=1.5431$, and $K_{\rm{t}} = 20$.

\subsubsection{Orbital instability}

The IL model describes the statistical outcome of gravitational interactions among planets with analytical plus Monte Carlo prescriptions for collisions and close scattering between planets \citep{Ida_2010,Ida_2013}, calibrated with results from $N$-body simulations \citep{Nagasawa_2008}.
Here we do not describe the details of the prescriptions in the IL model, as our investigation of the eccentricity distribution of giant planets mainly relies on direct $N$-body simulations (see Section \ref{subsec:N-body}). We only highlight the main idea of the treatment of giant planet interactions in the IL model. For more details of the prescriptions including interactions among small, terrestrial planets, we refer to \citet{Ida_2010} and \citet{Ida_2013}.
\begin{enumerate}
    \item After disk depletion, "giant planets" are identified by two conditions: (i) the mass is larger than 30 $M_\oplus$ (ii) $e_{\rm{esc}} = v_{\rm{esc}}/v_{\rm{K}} = \sqrt{2Gm/R_{\rm{p}}}/\sqrt{GM_*/a} > 1$, where $v_{\rm{esc}}$ is the escape velocity from the planet surface and $R_{\rm{p}}$ is the physical radius of the planet. 
    The orbit crossing time $\tau_{\rm{cross}}$ is evaluated for all giant planet pairs using the fitting formula given by \citet{Zhou_2007} with some modifications:
    \begin{equation}
        \log{\left( \frac{\tau_{\rm{cross}}}{T_{\rm{K}}}\right)} = A + B \log{\left(\frac{b}{2.3r_{\rm{H2}}}\right)},
        \label{eq:t_cross}
    \end{equation}
    where $T_{\rm{K}}$ is the Keplerian time at the mean orbital semi-major axis of the two planets $a = \sqrt{a_i a_j}$, $b = |a_i-a_j|$, $r_{\rm{H2}} = [(m_i+m_j)/3M_*]^{1/3} \min{(a_i, a_j)}$, and the coefficients are
    \begin{eqnarray}
        A &=& -2 + e' - 0.27 \log{\mu},  \nonumber\\
        B &=& 18.7 + 1.1 \log{\mu} - (16.8+1.2\log{\mu})e', \nonumber\\
        e' &=& \frac{1}{2}\frac{(e_i+e_j)a}{b}, \\
        \mu &=& \frac{(m_i+m_j)/2}{M_*}.  \nonumber
        \label{eq:t_cross_coefficients}
    \end{eqnarray}

    \item Orbit crossing is expected to occur at a time $t = \tau_{\rm{disk}} + \tau_{\rm{cross}}$. If the expected orbit crossing happens before the end of the simulation ($t<t_{\rm{end}}=10^9$ years), prescriptions of close encounters would be applied to the according giant planets. The pair with the shortest $\tau_{\rm{cross}}$ is assumed to undergo close encounters before any other pairs, because the system instability time is determined by the pair with the closest separation \citep{Yang_2023}. Other giants that participate in subsequent orbit crossing are selected based on whether their radial excursions overlap with the expected orbits of the first-crossing pair.
    New orbital elements and masses (in case of collisions) of the giant planets that undergo orbit crossing are calculated by the prescriptions (see Appendix in \citealt{Ida_2013} for details of the prescriptions).

    \item The orbit crossing time $\tau_{\rm{cross}}$ of all giant planet pairs are updated by the new orbital configuration of the system, and the time of the expected orbit crossing event $t$ is calculated again and compared with $t_{\rm{end}}$. These procedures are repeated until $t>t_{\rm{end}}$.

    \item All planets other than the giant planets are removed based on the assumption that violent secular perturbations from the highly-eccentric giant planets would destabilize the system and eject other planets \citep{Matsumura_2013}.
\end{enumerate}

% describe model setup in a table. 10 simulations in total.
\begin{table}
    \centering
    \renewcommand\arraystretch{1.2}
    \caption{Summary of simulations}
    \begin{tabular}{ccccc}
    \hline \hline
    \multicolumn{5}{c}{New model: PPS} \\\hline
    Model & $\Delta a$ (au) & $f_1$ & $N_{\rm{star}}$ & $m_{\rm{tot}}~(M_\oplus)$ \\\hline
    1 & 5 & 0.1 & 402 & 100 \\
    2 & 10 & 0.1 & 402 & 100 \\
    3 & 20 & 0.1 & 402 & 100 \\
    4 & 10 & 0.03 & 402 & 100 \\
    5 & 10 & 0.3 & 402 & 100 \\
    6 & 20 & 0.03 & 402 & 160 \\\hline\hline
    \multicolumn{5}{c}{New model: PPS$+N$-body} \\\hline
    7 & 10 & 0.1 & 100 & 100 \\
    8 & 10 & 0.03 & 100 & 100 \\
    9 & 20 & 0.1 & 100 & 160 \\
    10 & 20 & 0.03 & 100 & 160 \\\hline\hline
    \multicolumn{5}{c}{Classic model} \\\hline
    Model & $a_{\rm{min}}$, $a_{\rm{max}} ^{[1]}$ & $f_1$ & $N_{\rm{star}}$ & $\Sigma_{\rm{d}}~(\rm{g}~{cm}^{-2})$ \\\hline
    11 & 0.1, 20 & 0.1 & 402 & $10 f_{\rm{d}}\eta_{\rm{ice}} \left(\frac{a}{1~\rm{au}}\right)^{-1.5}$ \\\hline
    % 11 & 0.1, 20 & 0.1 & 402 & \multirow{2}*{$10 \eta_{\rm{ice}} \left(\frac{a}{1~\rm{au}}\right)^{-1.5}$} \\
    % 12 & 0.1, 20 & 0.1 & 100 &  \\\hline
    \end{tabular}
    \label{tab:model}
    \begin{tablenotes}
        \item[1]  [1]\ In the classic model, we set the minimum and maximum semi-major axis ($a_{\rm{min}}$, $a_{\rm{max}}$) of both the planetesimals and the embryos to be 0.1 and 20 au, since we mainly focus on CJ and planetesimal accretion beyond 20 au is very inefficient \citep{Ida_2018}.
    \end{tablenotes}
\end{table}

% 20241016 proof-reading until here
\subsection{Numerical simulations}  
Using the new model as described above, we run simulations of planet formation and examine the final distribution of planets resulting from different initial conditions.
We also run one simulation with the IL model (as in \citealt{Ida_2018}) and compare the results with those of the new model. 
The parameters of all models are summarized in Table \ref{tab:model}. 

\subsubsection{PPS simulation} \label{subsec:PPS_method}
Models 1--6 are PPS simulations (without $N$-body simulations) of the new model for the purpose of demonstrating the orbital distance and mass distribution (as in Section \ref{sec:a-m_distribution}). 

To ensure a fair comparison with observations, we use the mass and metallicity of the 402 stars in the RV sample as the initial conditions of the PPS simulations (see Figure \ref{fig:RV_mass_metallicity} in Appendix \ref{appendix:RV_star}). The distribution of the disk dispersal timescale $\tau_{\rm{disk}}$ is a log-normal function with a mean value of 2.5 Myr and a dispersion of 1. The timescale of the late-stage rapid disk clearing phase is $\tau_{\rm{rapid}} = 0.004 \tau_{\rm{disk}}$. The scaling factor for the gas surface density $f_{\rm{g}}$ is distributed in a range of [0.1, 10] with a log-normal function, which means that the gas surface density is 0.1-10 times the MMSN value.
The scaling factor for the solid surface density $f_{\rm{d}}$ also follows a log-normal distribution, with a mean value setting the total solid mass within the disk is $100~M_\oplus$ and a dispersion of 1. 

For the purpose of Section \ref{sec:a-m_distribution} and to investigate the dependence of the results on the disk width $\Delta a$ and the migration efficiency $f_1$, we design 5 models, each with 402 simulations (one simulation for one system/star). Model 2 is set as the fiducial model. The dependence on the disk width is explored through Models 1--3, and the dependence on the migration efficiency is explored through Models 2, 4, and 5.
Model 6 is introduced after the discussion of the eccentricity distribution of giant planets (see Section \ref{sec:eccentricity} and \ref{sec:discussion}).
Given the possible uncertainties of planetesimal formation, the width of the planetesimal disks in each model ($N_{\rm{star}}$ simulations) follows a normal distribution with a mean value of $\Delta a$ specified in Table \ref{tab:model}. In the models with a disk width of 20 au (Models 3, 6, 9, and 10), the dispersion of the disk width distribution is 2 au, while in other models, the dispersion is 1 au.  

Model 11 represents the IL model for comparison. The solid surface density in the IL model resembles that of the MMSN as $\Sigma_{\rm{d}} = 10 f_{\rm{d}} \eta_{\rm{ice}} (a/1~{\rm{au}})^{-1.5}$, where $\eta_{\rm{ice}}=1$ for $a<a_{\rm{ice}}$ and $\eta_{\rm{ice}}=4.2$ for $a\geq a_{\rm{ice}}$. The scaling factor $f_{\rm{d}}$ follows a log-normal distribution within the range of [0.1, 10]. 
% Similar to Models 1-6, we use 402 stars with mass and metallicity taken from the RV sample, and run PPS simulations for Model 11. The surface density (both gas and solid) of the disk is a power-law distribution. The disk mass is distributed within [0.1, 10] times the MMSN value with a log-normal distribution. 
The minimum and maximum semi-major axis for the planetesimals and planet seeds are set as 0.1 and 20 au, considering that we mainly focus on cold giant planets and that planetesimal accretion is very slow beyond 20 au \citep{Ida_2018}.

\subsubsection{PPS simulations integrated with N-body simulation}
\label{subsec:N-body}
The IL model uses a statistical approach to estimate the eccentricities of planets \citep{Ida_2010,Ida_2013}. This method has the advantage of producing relatively good eccentricity estimates for a large sample at low computational costs. In order to have more precise eccentricity estimates, we add $N$-body simulations of the planets after disk dispersal. 
Models 7--10 are PPS simulations integrated with $N$-body simulations for the purpose of demonstrating the eccentricity distribution of CJs. 
In both sections, we explore the dependence of the final distribution on $\Delta a$ and $f_1$. 
% describe how we added N-body simulations. How the initial conditions were intercepted from PPS.
To integrate the PPS simulation with $N$-body simulation, we list the planet data (including the mass $m_0$, the semi-major axis $a_0$, and the eccentricity $e_0$ of all planets larger than $0.1~M_\oplus$) at the time of disk depletion, and use these data as the initial conditions for orbital integration. 
The inclination is assumed to be half of the eccentricity ($i_0 = e_0/2)$). The argument of pericenter, the longitude of ascending node, and the time of pericenter passage are randomly chosen from 0 to $2\pi$. 
Then we use the REBOUND code \citep[e.g.,][]{Rein_2012} with the IAS15 integrator \citep{Rein_2015_IAS15} to calculate the orbital evolution of the planets in each system. Collisions happen whenever two particles physically touch each other and are resolved as mergers, conserving the mass, momentum, and volume. We remove a particle from the simulation when its distance from the central star exceeds $10^6$ au. All systems are integrated for 100 Myr (starting from the time of disk depletion). 

Due to the high computational cost of $N$-body simulations, we reduce the number of stars in each model from 402 to 100. To guarantee that we have the maximal size of the CJ sample, for each model we first run 402 PPS simulations as described in the previous sub-section. Then from these 402 simulations we choose 100 systems that produce giant planets while preserving the original distribution of the stellar mass and metallicity. 

Since the semi-major axis and mass distribution is mainly determined by the migration and accretion during the disk phase, orbital instability phase after disk dispersal does not significantly change the distribution in the semi-major axis and mass parameter space. Therefore, we use Models 7--10 for the purpose of examining the eccentricity distribution of CJs only.

\section{Orbital distance and mass distribution} \label{sec:a-m_distribution}
We first focus on the semi-major axis and mass distribution, as well as the occurrence rate of giant planets with respect to the orbital period. We compare the results from the IL model and the new model (fiducial), and then explore the dependence of the results on the disk width and the migration efficiency. 

\subsection{The IL model vs. the new model}

In this section, we compare the results from the IL model (Model 11) and the fiducial model (Model 2). 

\subsubsection{IL model}

\begin{figure}
    \centering
    \includegraphics[width=0.45\textwidth]{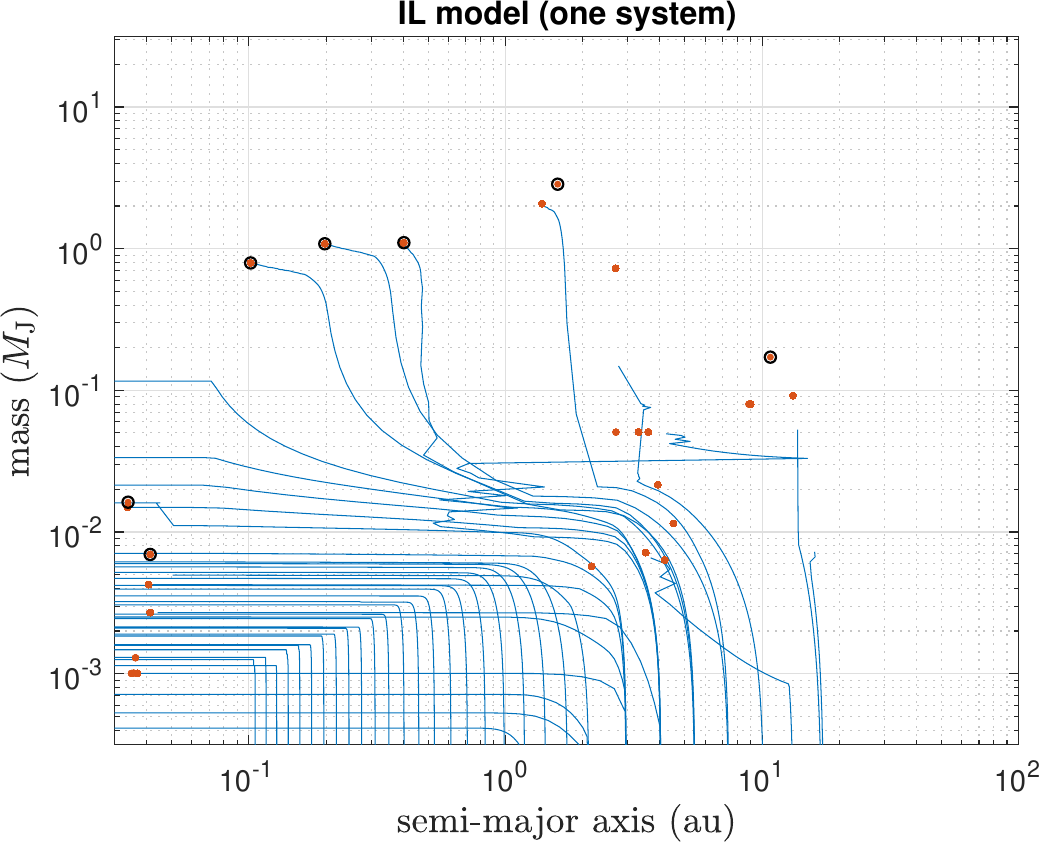}
    \caption{Growth tracks of planets on the $(a,m)$ diagram in one simulation (one system) using the IL PPS model. The blue lines show the evolution paths of planets during the disk phase. The red dots indicate the results of giant impacts after disk dispersal. The black open circles show the final surviving planets at the end of the simulation. The enhancement factor for gas surface density is set to $f_{\rm{g}} \simeq 5$ (5 times the MMSN value) to ensure that CJs can form in this system.}
    \label{fig:IL_growth_track}
\end{figure}

\begin{figure*}
    \centering
    \includegraphics[width=0.95\textwidth]{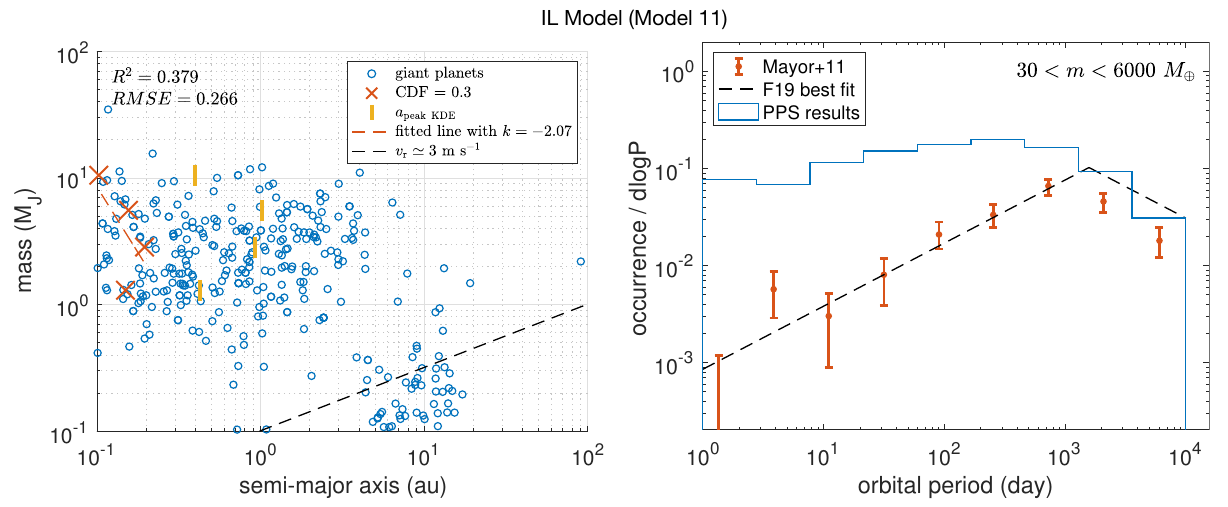}
    \caption{\textit{Left:} distribution of giant planets produced by the IL PPS model on the $(a,m)$ diagram (model 11). Similar to the left panel of Figure \ref{fig:RV_obs}, the inner boundary is indicated by the red dashed line. \textit{Right:} occurrence rate of giant planets within 30 to 6000 $M_\oplus$ as a function of orbital period. The blue histogram shows the results from the IL PPS model. Note that the planets below the RV detection limit line of $v_{\rm{r}} \simeq 3~\rm{m}~s^{-1}$ are excluded when calculating the occurrence rate because they are unlikely to be detected. The black dashed line represents the best fit given by the \texttt{epos} symmetric function parameters ($p_1 = 0.65$, $p_2 = -0.65$, $P_{\rm{break}} = 1581$ days) in \citet{Fernandes_2019} as in their Figure 4; the red dots with error bars are plotted with data extracted from their Figure 4 with permission.}
    \label{fig:IL_orbit_mass}
\end{figure*}

Figure \ref{fig:IL_growth_track} shows the growth tracks of planets in one simulation on the $(a,m)$ diagram. 
The end time of the simulation is $t_{\rm{end}}=10^9$ years.
This system is chosen from the simulations of Model 11 to show the typical planet growth tracks in a system in which giant planets can form (initially elevated gas surface density $f_{\rm{g}} \simeq 5$). 
The growth tracks show that giant planets typically form from cores that  grow beyond $a \simeq 3$ au, where the solid surface density is higher due to ice condensation. Most giant planets suffer from significant inward migration and become WJs. In this case, only two giant planets remain beyond 1 au, three become WJs, and one becomes a hot Jupiter (HJ) with $a<0.1$ au. %Such growth paths are also commonly seen in other works \citep[e.g.,][]{Ida_2013,Emsenhuber_2021_I}.
%Similar growth paths are also seen in \citet{Johansen_2019} (see their Figure 4).

\noindent \textit{- The inner boundary on the $(a,m)$ diagram}

The left panel in Figure \ref{fig:IL_orbit_mass} shows the distribution of giant planets produced by the IL model (Model 11) on the $(a,m)$ diagram. Since we focus on the density distribution on the $(a,m)$ diagram, the inner boundary where the KDE is a constant value (see description in Section \ref{sec:inner_boundary_obs}) is highlighted by the red dashed line in the left panel. From this scatter plot and the inner boundary, we can identify three features: (1) the inner boundary is unclear. No matter how the bin width is adjusted, the $R^2$ value cannot exceed 0.4, suggesting that there is no straight inner boundary as shown by the observation sample. (2) The location of the inner boundary is closer-in compared with that of the RV sample. %The outermost point of the inner boundary is within 0.3 au, while the innermost point of the inner boundary in the RV sample is beyond 0.3 au (see Figure \ref{fig:RV_obs}). 
(3) The fitted inner boundary has a negative slope, which is inconsistent with the inner boundary in the RV sample ($k \simeq 3$). 
This is mainly due to the fact that in the IL model where the solid material is uniformly distributed in the disk, migration generally outperforms accretion, so that many planets that managed to accrete enough gas to become giant planets suffer significant inward migration and land in the WJ zone. Similar results are also seen in \citet{Ida_2018}. While the important updates to the IL model introduced in \citet{Ida_2018} (the new Type-II migration formula and the two-$\alpha$ disk model) certainly solved the problem of the significant loss of giant planets due to migration and successfully retained a large proportion of cold giant planets, an inner boundary of giant planet distribution on the $(a,m)$ diagram that is consistent with the RV observation is not reproduced. 
    
\noindent \textit{- The pile-up of CJ near the snow line}

The right panel in Figure \ref{fig:IL_orbit_mass} shows the occurrence rate of giant planets (within the mass range of 0.1-20 $M_{\rm{J}}$) as a function of orbital period.
F19 points out that the occurrence of giant planets turns over at a location that corresponds to the snow line of Sun-like stars. %($P_{\rm{peak}} \simeq 1717$ days). 
It is clear from the comparison between the simulation results and the observation that in the results of Model 11, there is no prominent pile-up of giant planets near the snow line. In addition, the slope of the occurrence within the peak location $P_{\rm{peak}}$ is much flatter compared with observation. 
The reason is similar to what we have discussed above: the efficient inward migration of gas giants causes the over-abundance of WJs and the absence of a pile-up near 3 au. Although the occurrence of giant planets indeed decreases with decreasing orbital distance within $a \simeq 1$ au, the absolute abundance of giant planets within $a \simeq 1$ au is still too high compared with observation.

\subsubsection{The new model} \label{ssubsec:new_model}

\begin{figure}
    \centering
    \includegraphics[width=0.45\textwidth]{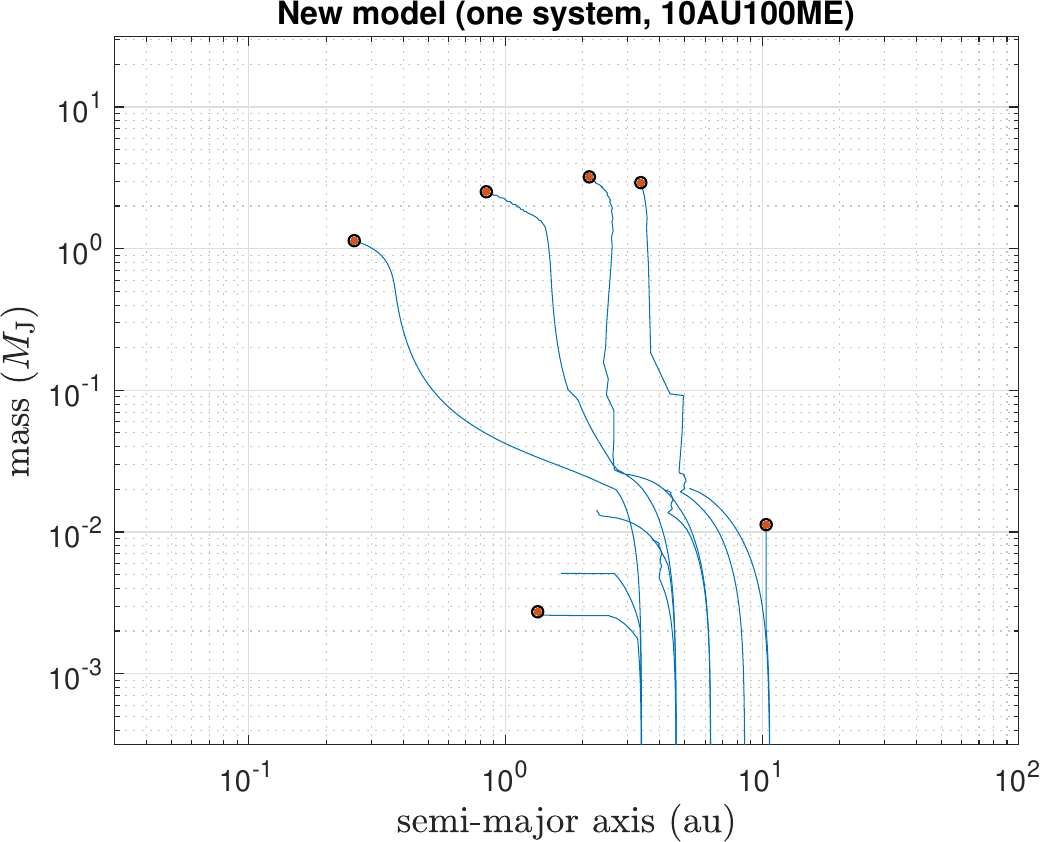}
    \caption{Same as Figure \ref{fig:IL_growth_track} but for a simulation using the new model. The initial planetesimal disk width is 10 au, and the total solid mass within the disk is 100 $M_\oplus$. The gas surface density enhancement factor $f_{\rm{g}}$ is set to 5 to guarantee gas giant formation for the purpose of demonstrating their growth paths. 
    }
    \label{fig:new_PPS_growth_path}
\end{figure}

\begin{figure*}
    \centering
    \includegraphics[width=0.95\textwidth]{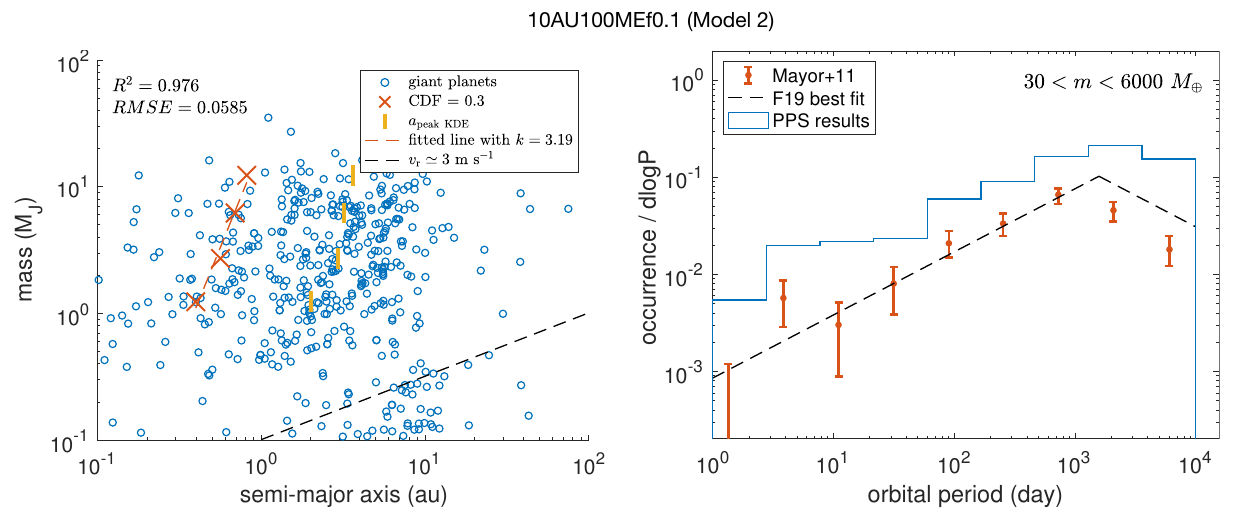}
    \caption{Same as Figure \ref{fig:IL_orbit_mass} but for Model 2 (new PPS model with truncated disks). The parameters for this model are listed in Table \ref{tab:model}.}
    \label{fig:model_2_orbit_mass}
\end{figure*}

We now show the results from the new model, in particular, the fiducial model (Model 2) with a moderate disk width of 10 au and the same migration efficiency as adopted in the IL model (see Table \ref{tab:model} for more parameters of this model). 
Figure \ref{fig:new_PPS_growth_path} shows the growth track of planets in one system chosen from Model 2. The system is also chosen with a gas surface density enhancement factor of $f_{\rm{g}} \simeq 5$ to ensure giant planet formation for the purpose of demonstrating their growth paths. Due to elevated solid surface density within the truncated planetesimal disk compared with the classic uniform disk model, the planet cores that grow within the disk can reach a gap-opening mass on a much shorter timescale and enter the Type-II regime where the migration is much slower. As a result, a large fraction of giant planets that grow beyond the snow line can stay beyond 1 au at the end of the simulation. In this case, two giant planets become WJs with one of them marginally crosses 1 au. The other two giant planets remain as CJs. 

\noindent \textit{- The inner boundary on the $(a,m)$ diagram}

The left panel in Figure \ref{fig:model_2_orbit_mass} shows the scatter plot of giant planets produced by the fiducial model (Model 2) on the $(a,m)$ diagram. %The inner boundary determined by the KDE is highlighted by the red dashed line. 
Compared with the results from the IL model, the results from Model 2 better reproduce the inner boundary of giant planet distribution on the $(a,m)$ diagram. Both the location and the slope of the inner boundary are consistent with that observed from the RV sample ($k \simeq 3$).
%The slope of the inner boundary ($k \simeq 2$) is also similar to that of the RV sample. 
As explained above, the elevated solid surface density within the disk truncated at the snow line reduces the time of core growth so that the growing planets can avoid undergoing too much efficient Type-I migration. Therefore, the majority of giant planets remain not far from the location where they were born. 

\noindent \textit{- The pile-up of CJ near the snow line}

The right panel in Figure \ref{fig:model_2_orbit_mass} shows the occurrence rate of giant planets as a function of orbital period. %Again, the blue histogram shows the simulation results, and the red dots and the black dashed line are taken from F19. 
Unlike in the IL model, in Model 2, the giant planet occurrence shows a peak within the bin of [1292-3594] days, which corresponds to [2.32-4.59] au for $M_* = 1~M_\odot$. Such a pile-up of giant planets is consistent with both the results reported in F19 and \citet{Fulton_2021}. Moreover, the slope of the occurrence on the left-side of the pile-up is similar to that of the power-law fit given in F19.
Since in the truncated disk model, the highest solid surface density appears at the snow line (see Equation \ref{eq:Sigma_d}), giant planets are most likely born near the location of the snow line. Due to the high efficiency of core growth, the accreting proto-giants avoid significant inward migration and finally stay not far from the snow line. Consequently, the occurrence of giant planets reaches the maximum near the snow line of Sun-like stars, since the given stellar mass peaks at $M_* \simeq 1~M_\odot$.

Our new models appear to over-estimate the occurrence of giant planets beyond the snow line (outside the pile-up location) compared with the observation. This is likely an artifact of the observation bias that planets on distant orbits are harder to be detected. Although we have already removed the planets below the $v_{\rm{r}} \simeq 3~\rm{m}~s^{-1}$ line when calculating the occurrence rates, the values of the simulation results are still a few factors larger than the observation (see also results of other models in Section \ref{subsec:dependence}). Since the error bars of the observationally inferred occurrence rates at distant orbits can be quite large (see also \citealt{Fulton_2021}), we do not intend to compare the absolute occurrence rate of the simulation results with the observation, and focus on the presence and the location of the peak instead.

\subsection{Dependence on the disk width and migration rate} \label{subsec:dependence}

In this section, we explore the dependence of the results, particularly the inner boundary on the $(a,m)$ diagram and the occurrence rate along the orbital period, on two important parameters, i.e., the width of the disk $\Delta a$ and the migration efficiency, represented by the retardation factor $f_1$. 

\subsubsection{Dependence on the disk width}

\begin{figure*}
    \centering
    \includegraphics[width=0.98\textwidth]{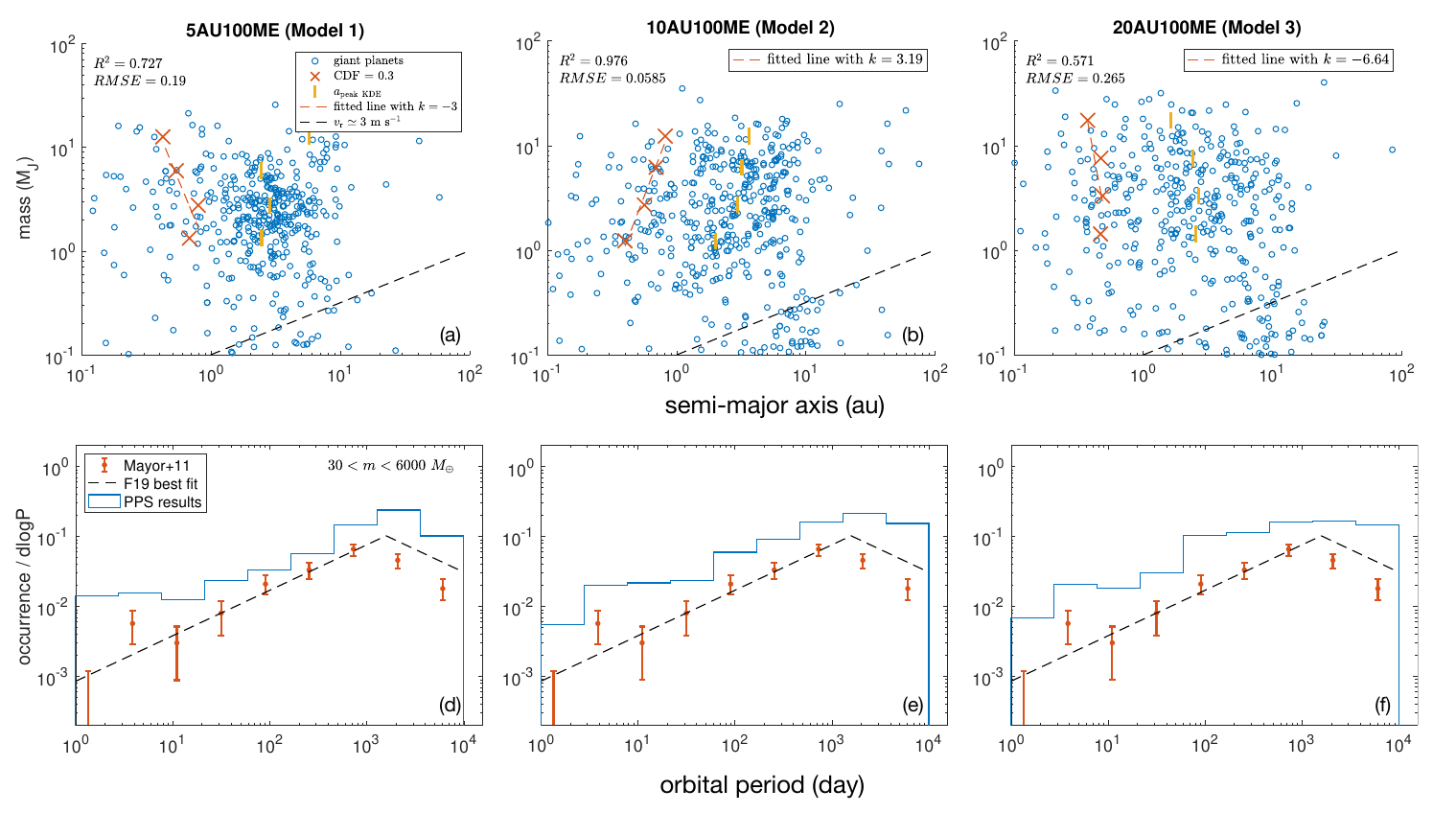}
    \caption{Comparison of the results from three models, each with a different disk width $\Delta a$ and all the other parameters remain the same. The parameters of each model are listed in Table \ref{tab:model}. \textit{Panels (a)--(c):} distribution of giant planets and the inner boundary on the $(a,m)$ diagram in Models 1--3. \textit{Panels (c)--(d):} occurrence rate of planets within 30 to 6000 $M_\oplus$ as a function of orbital period in Models 1--3.}
    \label{fig:disk_width_6_plots}
\end{figure*}

Figure \ref{fig:disk_width_6_plots} shows the results from Models 1--3, each with a different disk width $\Delta a$ but the total solid mass is fixed as $m_{\rm{tot}} \simeq 100~M_\oplus$.
The upper row compares the distribution of giant planets on the $(a,m)$ diagram in the three models. The inner boundary location barely depends on the width of the disk. In all three cases, the inner boundary location barely varies. This is a natural result since no matter how the disk width changes, the inner edge of the disk remains at the location of the snow line. However, the slope of the inner boundary sensitively depends on the disk width: taking panel (b) as the standard, the slope becomes much steeper (reaching a negative value) when the disk is narrow (panel a, Model 1), and slightly steeper when the disk becomes wider (panel c, Model 3). When the disks are narrow, the distribution of planets on the $(a,m)$ diagram is rather concentrated. In other words, the results of planet formation are similar (with similar mass and orbital distance) when the given initial disk width is narrow. %In this case, the linear fit does not describe the inner boundary very well (this is shown by the lower $R^2$ value of the fit in panel a compared with those in panels b and c). 
When the disks are wider, the solid surface density at a certain location is lower compared with the narrower-disk case, because the total mass of solids is fixed. Consequently, fewer giant planets can form and migration becomes more significant, making the planets more dispersed in the $(a,m)$ diagram. As the disk becomes wider, the scatter plot on the $(a,m)$ diagram as well as the shape of the inner boundary becomes more similar to that of the results from the IL model. 

The bottom row compares the occurrence rate of giant planets with respect to the orbital period. It is clear that a peak of occurrence rate at the period bin of [1292-3594] days exists in all three models. This suggests that the peak occurrence location does not depend on the width of the disk. The reason is simple: the inner edge of the truncated disk remains at the location of the snow line, and most of the giant planets that form within truncated disks are able to stay not far from the snow line (see Section \ref{ssubsec:new_model}). Therefore, as long as the inner edge of the disk does not change, the peak location of the occurrence would not significantly vary. However, the slope of the occurrence within the pile-up location ($P_{\rm{peak}}$) slightly flattens as the disk becomes wider. Giant planets undergo more efficient migration because the local solid surface density is lower in the wider disk compared with the narrower disk, so that growing planets take longer time to reach the gap-opening mass. As a result, the fraction of WJ is higher in the wider disks.

\subsection{Dependence on the migration rate} 

\begin{figure*}
    \centering
    \includegraphics[width=0.98\textwidth]{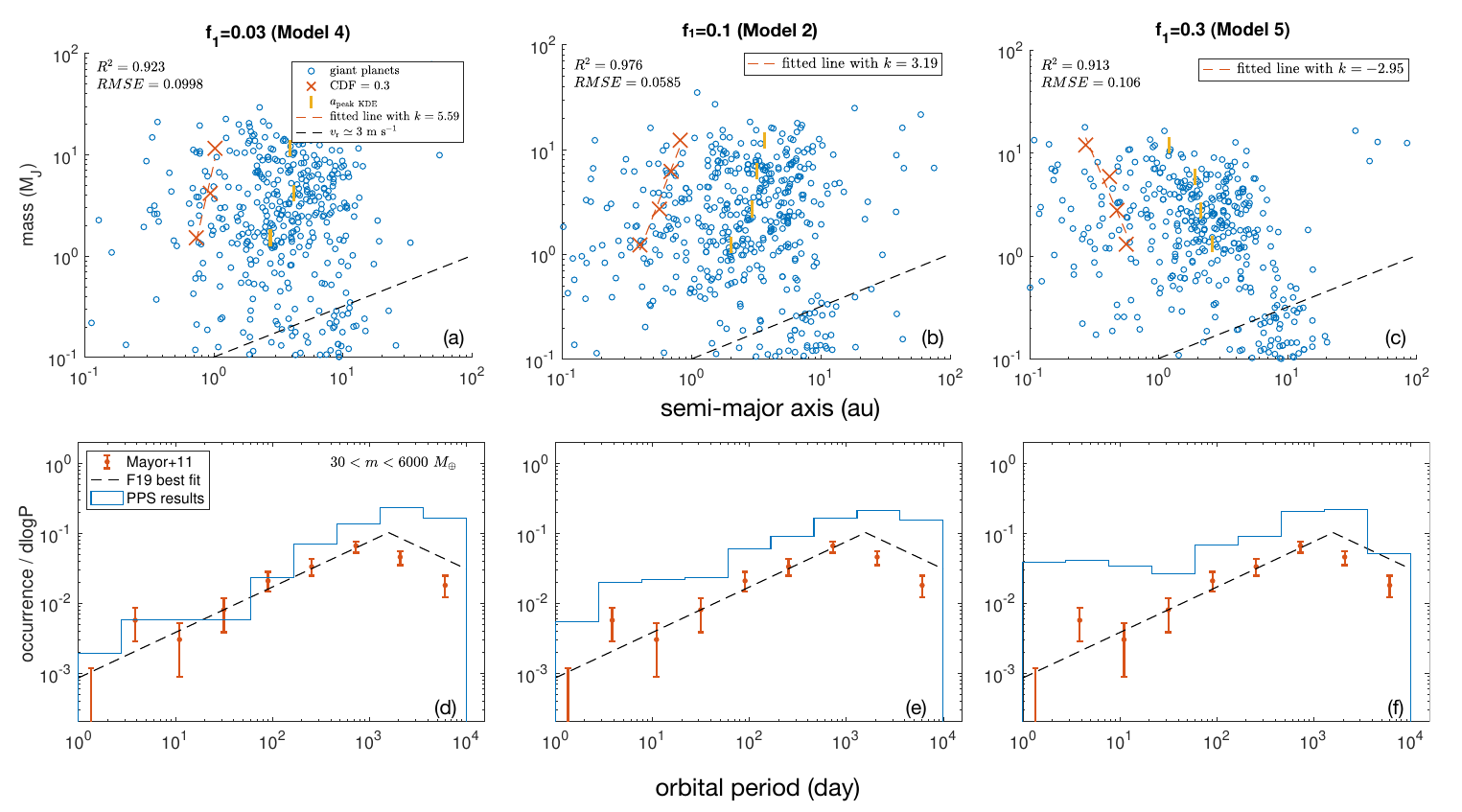}
    \caption{Comparison of the results from three models, each with a different migration retardation factor $f_1$ and all the other parameters remain the same. The parameters of each model are listed in Table \ref{tab:model}. Same as Figure \ref{fig:disk_width_6_plots}.}
    \label{fig:migration_6_plots}
\end{figure*}

Figure \ref{fig:migration_6_plots} shows the dependence of the results on the efficiency of migration $f_1$, with a focus on the $(a,m)$ diagram distribution and the occurrence rate along orbital period. 
Similar to Figure \ref{fig:disk_width_6_plots}, the upper row shows the scatter plot of giant planets as well as the inner boundary on the $(a,m)$ diagram in Models 4, 2, and 5. Comparing panels (a)--(c), it is easy to see that the inner boundary moves inward as the migration rate increases. %The location of the innermost point of the inner boundary moves from $\simeq 0.7$ au (panel a) to $< 0.3$ au (panel c), as the retardation factor of migration increases from 0.03 to 0.3. 
This is a natural outcome of the giant planets moving inward due to faster migration. As for the shape of the inner boundary, the slope becomes slightly steeper when migration is slow, and becomes negative when migration is fast. In the slow-migration case ($f_1=0.03$), the distribution of giant planets is more confined along the $a$-axis, causing the distribution of data points to be more vertical in the $(a,m)$ space, so that the inner boundary becomes steeper. In the fast-migration case, the shape of the inner boundary is more similar to that of the results from the IL model (see Figure \ref{fig:IL_orbit_mass}). The massive planets can also undergo significant inward migration even after gap opening, resulting in a negative slope. 

The bottom row demonstrates the dependence of the giant planet occurrence rate along the orbital period on the migration efficiency. Again, the peak location of the occurrence barely changes in these three models, meaning that the pile-up of giant planets is not affected by the migration rate in the truncated disk model. The slope of the occurrence on the left-side of the peak becomes flatter as the migration rate increases, owing to the formation of more WJs.

\section{Eccentricity distribution} \label{sec:eccentricity}

\begin{figure*}
    \centering
    \includegraphics[width=0.9\textwidth]{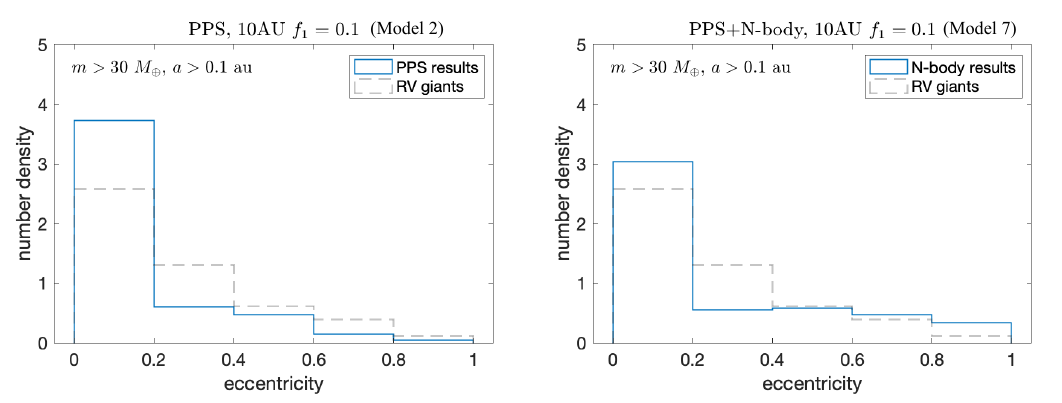}
    \caption{Eccentricity distribution of giant planets with masses larger than 30 $M_\oplus$ and semi-major axis larger than 0.1 au. \textit{Left:} results from Model 2 (10-au disk case with the new PPS model) with no $N$-body simulation implemented. \textit{Right:} results from Model 7 (10-au disk case with $N$-body simulations integrated with the new PPS model. The blue histogram shows the simulation results, and the grey dashed line shows the eccentricity distribution of planets in the RV sample.}
    \label{fig:ecc_histo_wo_Nbody}
\end{figure*}

\begin{figure*}
    \centering
    \includegraphics[width=0.9\textwidth]{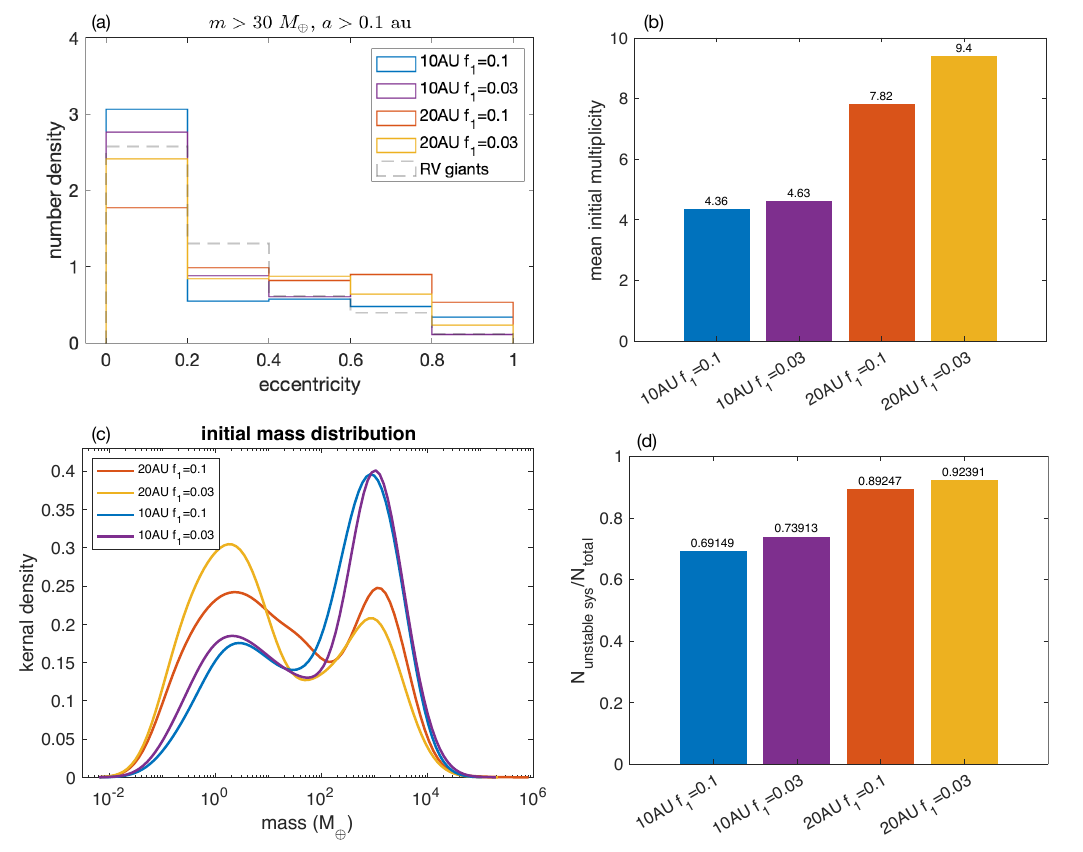}
    \caption{(a) Same as Figure \ref{fig:ecc_histo_wo_Nbody} but for all four models from Model 7-10. (b) The mean initial multiplicity of planets in each model (which includes 100 systems). (c) The kernel density distribution of the mass of all planets in each model. (d) The fraction of unstable systems ($N_{\rm{ustable}}/N_{\rm{total}}$) in each model. The fraction is calculated by dividing the number of unstable systems $N_{\rm{unstable}}$ by the total number of systems ($N_{\rm{total}}=N_{\rm{unstable}}+N_{\rm{stable}}$). The models are represented by different colors.}
    \label{fig:eccentricity_4_plots}
\end{figure*}

The eccentricity of giant planets reveals crucial information on their dynamical evolution history. In the previous section, we focused on the orbital distance and mass distribution of the giant planets, which is determined by the growth and migration history of planets in the disk phase through disk-planet interaction. After disk dissipation, the gravitational interaction between giant planets becomes important and determines the dynamical evolution and final orbital architecture of the system.
In this section, we shift our emphasis to the eccentricity distribution of giant planets (especially CJ) and compare our simulation results with observation.

First we compare the results from the PPS simulations and $N$-body simulations.
Figure \ref{fig:ecc_histo_wo_Nbody} shows the eccentricity distribution of giant planets with mass larger than 30 $M_\oplus$ and semi-major axis larger than 0.1 au in the fiducial model. The left panel shows the results from the PPS simulation only (Model 2) and the right panel shows the results of the model where $N$-body simulation is integrated to the PPS simulation (Model 7). In both panels, the grey histogram shows the eccentricity distribution of the giant planets in the RV sample. 
%The left panel shows that PPS simulations alone cannot reproduce the observed eccentricity distribution of giant planets because the fraction of low-$e$ planets ($e<0.2$) is too high compared with observation. 
Comparing the left and the right panels, we observe that while the $N$-body simulations and the semi-analytical planet scattering model in the original PPS simulations produce results that are qualitatively consistent with each other, the $N$-body simulations improve the eccentricity distribution of giant planets by decreasing the fraction of planets with $e<0.2$ because they include secular perturbations to pump up the eccentricities, which is not implemented in the PPS model.
In the mean time, in both PPS and $N$-body simulation results, there is a deficit of giant planets with moderate eccentricity ($0.2<e<0.4$). This deficit of moderate-$e$ planets barely depends on the choices of parameters.

We further explore the eccentricity distribution of giant planets by varying the disk parameters including the disk width $\Delta a$ and migration efficiency $f_1$. Note that unlike in Models 1--5 where we always use the same total solid mass $m_{\rm{tot}}$, in Models 7--10 when we change the disk width $\Delta a$, we also change the total solid mass $m_{\rm{tot}}$ by fixing the solid surface density. This is because the dynamical instability of a system is significantly affected by the multiplicity of giant planets. Therefore, we quantitatively explore the dependence of the eccentricity distribution on the giant planet multiplicity by varying the disk width while maintaining the solid surface density. 

Figure \ref{fig:eccentricity_4_plots} shows the results from 4 models with $N$-body simulations (Models 7--10). 
Panel (a) is similar to Figure \ref{fig:ecc_histo_wo_Nbody} and shows the eccentricity distribution of giant planets in all four models. 
Panel (b) shows the mean initial multiplicity of planets in the 4 models at the beginning of the $N$-body simulation (i.e., at the end of the disk phase). The mean is the average value for the 100 systems in each simulation.
Panel (c) shows the kernel density of the planet mass distribution in each model at the beginning of the $N$-body simulation (i.e., at the end of the disk phase).
Panel (d) shows the fraction of unstable systems (the number of unstable systems divided by the total number of systems) in each model.
Comparing the blue and the red histogram (or purple and the yellow histogram) in panel (a), we see that increasing the disk width (while keeping the surface density) lowers the fraction of low-$e$ planets. This is because the giant planet multiplicity at the end of the disk phase is higher when the disk is wider, as shown in panel (b). In disks with a width of $\Delta a \simeq 10$ au (blue and purple histogram), the change in the migration rate barely affects the eccentricity distribution of giant planets. However, in the 20-au disks, slower migration produces an eccentricity distribution which is more consistent with observation (more $e<0.2$ planets and fewer $e>0.4$ planets in the yellow histogram compared with the red histogram). This is a consequence of the different mass distribution at the end of the disk phase for the two cases. As shown in panel (c), in the 20-au disk case, $f_1=0.03$ (slower migration) model produces more lower-mass planets and fewer giant planets, because the distant embryos beyond $a \sim 10$ au have long growth timescale and cannot accrete additional material within their orbits or collide with other embryos due to limited radial excursions.
These lower-mass planets can easily be ejected or scattered by the giant planets, resulting in a slightly higher fraction of unstable systems (compare red and yellow bars in panel d). However, the ejection or scattering of these lower-mass planets would not significantly increase the eccentricity of the giant planets, resulting in a higher fraction of low-$e$ giant planets and fewer high-$e$ giant planets compared with the $f_1=0.1$ (nominal migration) model. In the 10-au disk case, the initial mass distribution is almost the same even when the migration rate is different, so that the final giant planet eccentricity distribution is barely affected. 
Comparing the 4 models with the RV sample, we find that when the fraction of unstable systems is higher than $\simeq 70$\%, the eccentricity distribution of giant planets beyond 0.1 au is reasonably consistent with observation, although the simulations generally underestimate the number of planets with $0.2<e<0.4$.

\section{Discussion} \label{sec:discussion}

\subsection{Which models are consistent with CJ observation?}
\label{subsec:best model}

\begin{figure*}
    \centering
    \includegraphics[width=0.98\textwidth]{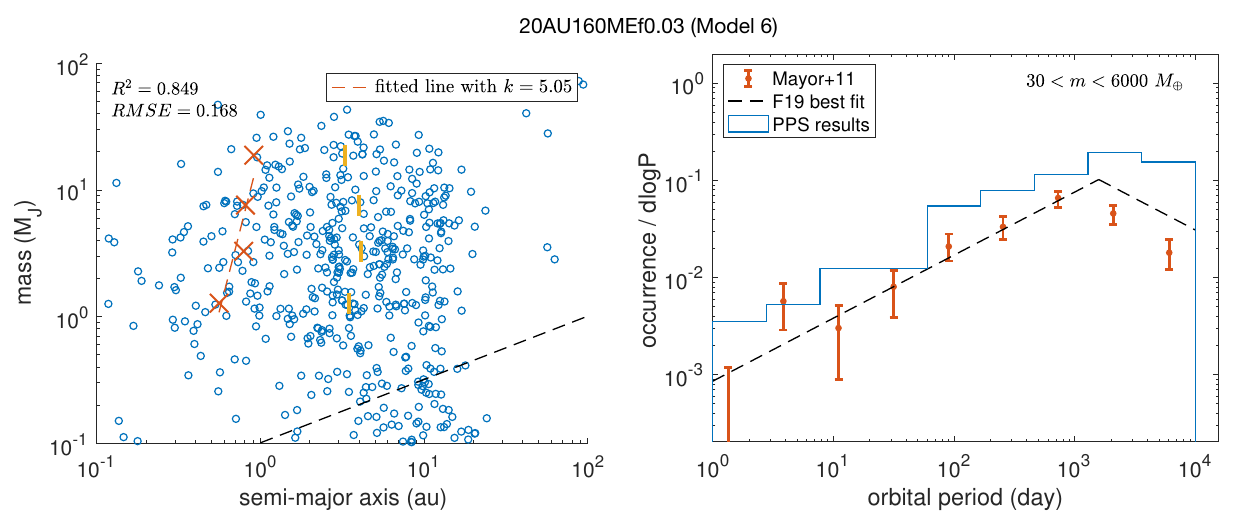}
    \caption{Same as Figure \ref{fig:IL_orbit_mass} but for the results of Model 6 ($\Delta a = 20$ au, $f_1 = 0.03$, $m_{\rm{tot}}\simeq 160~M_\oplus$). $N$-body simulation is not included in this model.}
    \label{fig:model_6_orbit_mass}
\end{figure*}

In Section \ref{sec:a-m_distribution}, we compared the results from five different models (1--5), with a focus on the distribution of giant planets and the inner boundary on the $(a,m)$ diagram, and the occurrence rate of giant planets along their orbital period. In Section \ref{sec:eccentricity}, we shifted our focus to the eccentricity of giant planets and compared the results of four models (7--10). The width of the planetesimal disk and the migration efficiency were two important parameters we explored throughout these models. 
Looking at the results from those two sections, we find that a relatively wide disk ($\Delta a \gtrsim 10~$au) with a suppressed migration rate ($f_1 \lesssim 0.1$) would produce a giant planet population with orbital and mass distribution that is reasonably consistent with the RV sample. 
This includes models 2, 4, and 10. 

Since in Section \ref{sec:a-m_distribution}, we explored the disk width dependence by varying the disk width and keeping the total solid mass, the parameter set in model 10 (which we performed $N$-body simulations in Section \ref{sec:eccentricity}) was not investigated to show the resulting giant planet distribution on the $(a,m)$ diagram and the occurrence rate along orbital period. Therefore, we add model 6 to perform a large-sample PPS simulation as in Section \ref{sec:a-m_distribution} to check the inner boundary distribution and pile-up location of giant planets. In this model, we use the same disk width and migration efficiency as in model 10, but instead of giving only 100 stars, we use the same stellar sample as in the RV sample (402 stars). 
Figure \ref{fig:model_6_orbit_mass} shows the distribution of giant planets on the $(a,m)$ diagram as well as the occurrence rate of giant planets along the orbital period, resulting from model 6. The left panel shows that, as expected from Section \ref{subsec:dependence}, the less efficient migration makes the inner boundary slightly steeper (a larger $k$ but still positive slope) compared with the fiducial model (2). The larger disk width also contributes to increasing $k$ as in model 3, but less significantly since the local surface density is fixed when increasing the disk width. The location of the inner boundary is also qualitatively consistent with the observation.%, with the innermost point located beyond 0.3 au. 
The right panel shows that the occurrence of giant planets along the orbital period is in good agreement with observation: the pile-up of giants appears near the location of the snow line, and the slope of occurrence within the pile-up is consistent with the observation.

In summary, models 2, 4, and 10 reproduce the features of the RV sample reasonably well. 

\subsection{What determines the inner boundary on the (a,m) diagram?}

\begin{figure*}
    \centering
    \includegraphics[width=0.98\textwidth]{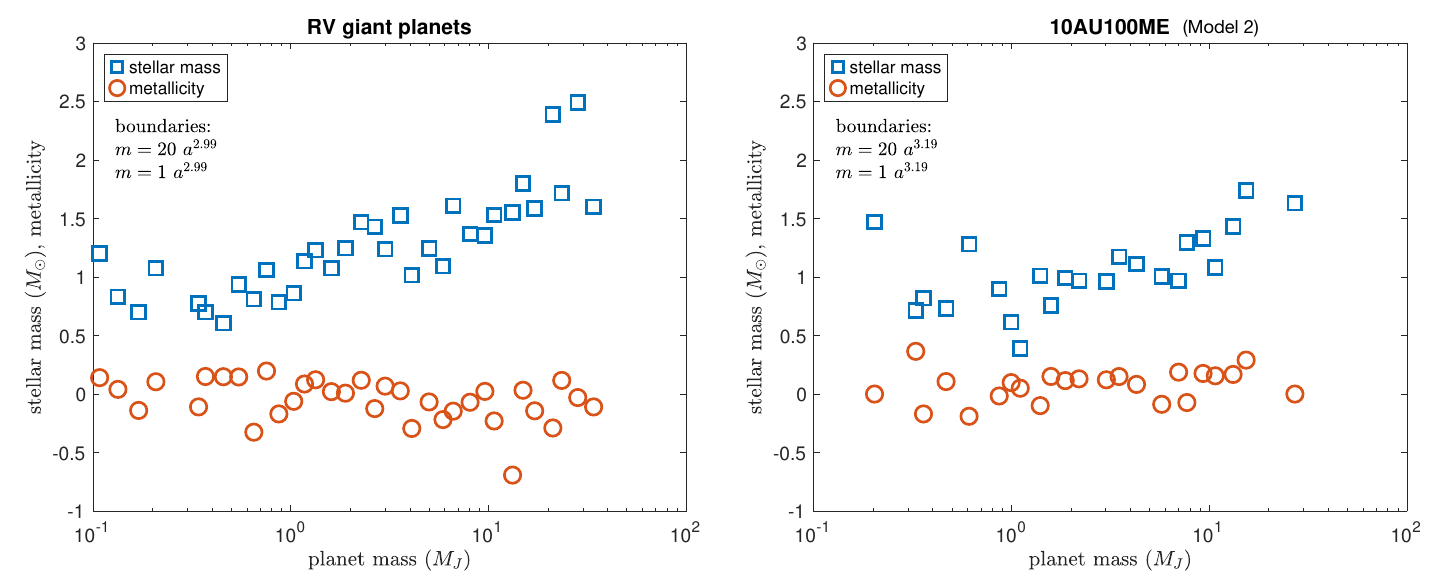}
    \caption{Correlation between the planets along the inner boundary on the $(a,m)$ diagram and the stellar mass and metallicity. Planets are chosen along the inner boundary between two lines with a slope of $k$. The $x$-axis is the planet mass which is correlated with the semi-major axis by $m \propto a^k$ (or $m\sin{i} \propto a^k$ in the RV sample). The left panel shows the RV sample, and the right panel shows the results of our fiducial model (Model 2). The $k$ values are taken from Figure \ref{fig:RV_obs} and Figure \ref{fig:model_2_orbit_mass} respectively.}
    \label{fig:correlation}
\end{figure*}

A key assumption in our models is that the planetesimal disk is truncated at the snow line. This allows our model to accurately reproduce the observed inner boundary of giant planet distributions on the $(a,m)$ diagram and the concentration of giant planets near the snow line. As discussed in Section \ref{ssubsec:new_model}, in a truncated disk with locally enhanced solid surface density, the core grows rapidly, enabling proto-gas giants to avoid significant inward migration and remain close to their birth locations. Since solid surface density peaks at the snow line, the occurrence of giant planets is maximized there.
This indicates that, on the $(a,m)$ diagram, planets along the inner boundary are correlated with stellar mass, as the snow line's location depends on stellar mass. Figure \ref{fig:correlation} illustrates this correlation between planet mass, semi-major axis, and stellar properties (mass and metallicity). The left panel presents the analysis for the RV sample, while the right panel shows the analysis for the results of the fiducial model. To identify planets along the inner boundary, we draw two lines with the same slope $k$ in the $(a,m)$ diagram ($m = C_1 a^k$ and $m = C_2 a^k$), selecting coefficients $C_1=1$ and $C_2=20$ to ensure a sufficient number of planets between the lines. We categorize planet mass into 40 logarithmically spaced bins from 0.1 to 50 $M_{\rm{J}}$, correlating their mass and semi-major axis as $m \propto a^k$. We then calculate the mean stellar mass and metallicity for each bin, plotting these values against the semi-major axis. In the left panel, planet mass is represented by the minimum mass $m\sin{i}$ for the RV sample.
%\textbf{We use the RV sample to represent the observations here, as it offers a larger sample size and a broader range of stellar mass and metallicity, allowing trends to be more clearly identified. Moreover, planets along the inner boundary on the $(a, m)$ diagram are far less affected by detection bias compared to those on more distant orbits. This reduces the need to rely on a homogeneous sample with controlled bias.}

Comparing the left and right panels reveals that planets along the inner boundary in the $(a,m)$ diagram show increasing stellar mass with increasing planet mass and semi-major axis, while metallicity remains relatively constant. This underscores the significant role of the snow line's location—and therefore stellar mass—in shaping the inner boundary of giant planet distributions on the $(a,m)$ diagram.
We note that by adopting the location of the snow line in an optically thin disk which is determined by stellar irradiation, our results suggest that CJs typically form in optically thin disks when the snow line is located at a few au. 

\subsection{Relationship between outer giant companions and inner small planets}
\label{subsec:relation}

In our truncated disk model, we focus on the population of giant planets, particularly cold Jupiters (CJs), by considering only planetesimal disks beyond the snow line. This approach entirely excludes the formation of hot/warm giant planets and inner small planets in situ. However, inner small planets and close-in giant planets can still form through sufficient inward migration under certain conditions. Thus, it is necessary to examine the relationship between inner small planets and outer giant planets and compare this with observations.

Using the CLS sample, \citet{Rosenthal_2022} found that stars hosting both inner small planets (0.02-1 au and 2-30 $M_\oplus$) and outer companions (0.23-10 au and 3-6000 $M_\oplus$) are significantly more metal-rich than those hosting only inner small planets. Additionally, the outer giant companions of inner small planets typically exhibit lower eccentricities compared to giant planets without inner companions (referred to as lonely giants).

In our simulation results, we observe a similar difference in metallicity distribution between stars with only inner small planets and those hosting both outer giant companions and inner small planets, consistent with \citet{Rosenthal_2022}. We also compare the eccentricities of outer giant companions to inner small planets with those of lonely giant planets and find that the former have lower eccentricities, aligning with the observations of \citet{Rosenthal_2022}.

We interpret these results as follows: stars that host lonely giants (those without inner small planets) tend to be more metal-rich, leading to the formation of more massive giant planets, which results in greater dynamical instability and higher eccentricities. Furthermore, lonely giant planets generally exhibit higher multiplicity compared to the outer companions that coexist with inner small planets, attributed to the higher metallicity of their host stars and the greater availability of solid mass in the disk. In our fiducial model 2, the multiplicity distribution of the lonely giants peaks at 2 and has a long tail extending up to 8, while the outer companions show a peak multiplicity of 1 and a shorter tail extending to 5. Consequently, this increases the likelihood of instability, producing more high-eccentricity giant planets in systems of lonely giant planets.

These findings suggest that locally confined planetesimal distributions with enhanced solid surface density—especially those truncated at the snow line—are ideal environments for the formation of CJs.

\section{Conclusions} \label{sec:conclusion}

Inspired by the latest observation of CJs (mainly detected by RV) and theories of planet formation from discrete planetesimal rings, we propose a new model of CJ formation from planetesimal disks that are truncated at the water snow line. Focusing on the distribution of giant planets on the $(a,m)$ diagram, the occurrence rate of giant planets along the orbital period, and the eccentricity distribution, we compare the simulation results of our models with the observation sample and explore the dependence of the results on the disk width and migration efficiency. 

We find that compared with the original IL model, the truncated disk model can better produce the observed inner boundary of giant planet distribution on the $(a,m)$ diagram, which shapes the "desert" (low occurrence region) of WJs. Classical models using uniform solid surface density distribution produce inner boundaries too close-in with a negative slope, while the observation shows that the inner boundary is located beyond 0.3 au and has a positive slope of roughly 2.
We also find that, while the location of the inner boundary barely depends on the width of the truncated disk, it moves inward as the migration efficiency increases. In addition, the slope of the inner boundary is sensitive to the competition of growth and migration, which is affected by both the disk width and the migration rate.

The pile-up of the giant planets (maximum occurrence rate) near the snow line can also be naturally explained in the truncated disk model. While the results of classical models typically show no prominent pile-up of giant planets near the snow line, the truncated disk models generally produce a clear peak of giant planet occurrence near the snow line. The location of the peak barely depends on the disk width or the migration rate.
The slope of the occurrence within (leftwards of) the peak location almost does not depend on the disk width but flattens as migration becomes more efficient.

Looking at the eccentricity distribution of giant planets, we find that PPS simulations generally over-estimate the fraction of low-eccentricity ($e<0.2$) giant planets compared with observation. Adding $N$-body simulations at the end of the disk phase improves the eccentricity distribution by enhancing the fraction of planets with higher eccentricities ($e>0.2$). 
When the fraction of unstable systems is increased to over $\simeq 70$\% (by increasing the disk width or reducing the migration rate), the eccentricity distribution of giant planets is reasonably consistent with observation, except for the deficit of planets with $0.2<e<0.4$.

In conclusion, we find that a relatively wide disk ($\Delta a \gtrsim 10$ au) with suppressed migration ($f_1 \lesssim 0.1$) would produce a giant planet population with orbital and mass distribution that is reasonably consistent with RV observation.

\vspace{12pt}
\noindent \textit{Acknowledgements}
\vspace{6pt}

We thank Antoine Petit, Max Goldberg, Dong Lai, Doug Lin, Yifan Xuan, and Guangyao Xiao for valuable comments and useful discussions. This work is supported by the National Natural Science Foundation of China (Nos. 12250610186, 12273023).
Y. H. is supported by JSPS KAKENHI Grant Number JP24H00017. We thank the anonymous reviewer for valuable suggestions to help us improve the quality of this paper. 
We thank Benjamin J. Fulton for sharing the detection completeness data for the CLS planet sample.
Permissions from Rachel B. Fernandes were acquired for using extracted data from F19 in figures in this manuscript.
The numerical computations of models including $N$-body simulations (Models 7--10) were conducted on the general-purpose PC clusters at the Center for Computational Astrophysics, National Astronomical Observatory of Japan. 

\appendix

\section{Reliability of the RV sample} \label{appendix:RV_reliability}

We used a composite RV sample acquired from the NASA Exoplanet Archive as the observation sample to be compared against our theoretical model results. Although this composite RV sample is larger and more modern, it suffers from inhomogeneity, as the planets were detected using different instruments and surveys. This inhomogeneity raises concerns about the reliability of certain demographic features, particularly the inner boundary and eccentricity distribution, which are central to this study.
%Our analysis of the RV sample relies solely on the detection efficiency reported in \citet{Mayor_2011}.}

To evaluate whether our analysis of the composite RV sample adequately captures the true features of giant planet demographics, we use two more robust samples with controlled biases, i.e., the planet sample from \citet{Mayor_2011} and the California Legacy Survey (CLS, \citealt{Fulton_2021}), to perform the same calculation of the inner boundary on the $(a, m)$ diagram as described in Section \ref{sec:inner_boundary_obs}. The eccentricity distributions are also plotted for comparison. 

\begin{figure*}
    \centering
    \includegraphics[width=0.98\textwidth]{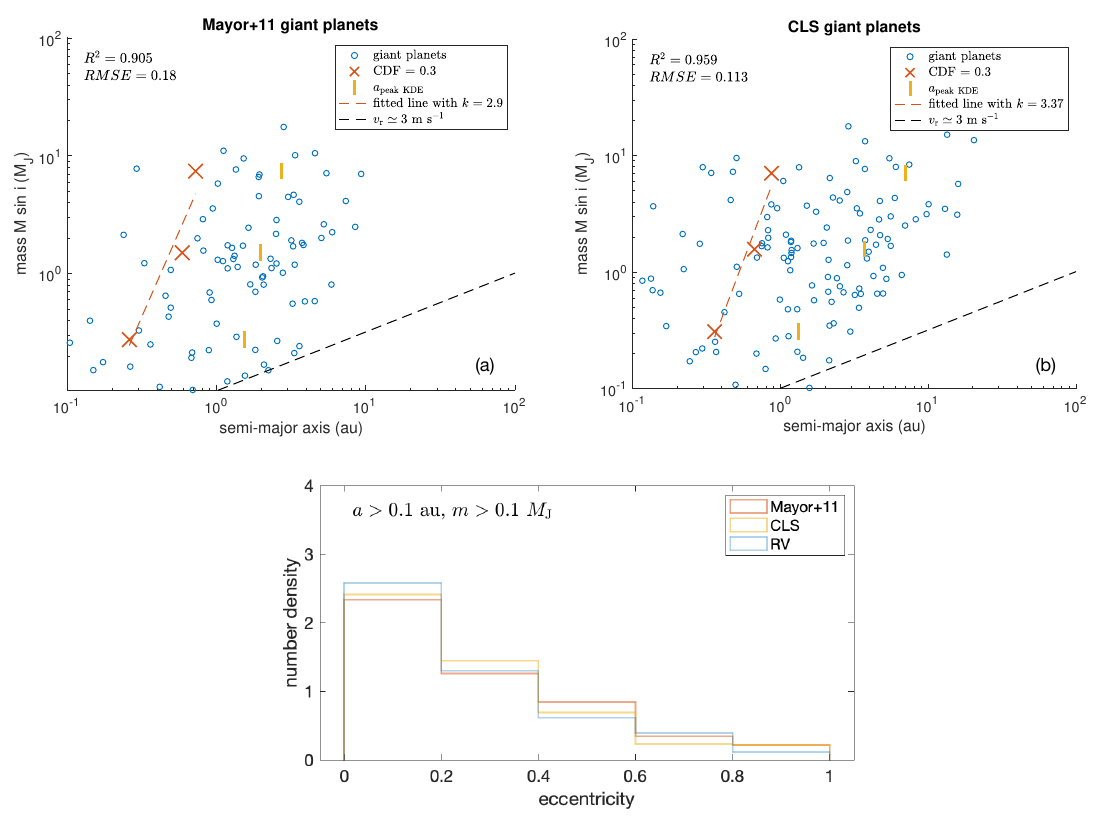}
    \caption{\textit{Panels} (a)-(b): the inner boundary of giant planet distribution on the $(a,m)$ diagram in the \citet{Mayor_2011} and CLS sample, respectively. \textit{Panel} (c): comparison of the eccentricity distribution in the RV sample, the \citet{Mayor_2011} sample, and the CLS sample.}
    \label{fig:compare_3_samples}
\end{figure*}

Panels (a) and (b) of Figure \ref{fig:compare_3_samples} display the distributions of the \citet{Mayor_2011} and CLS samples on the $(a, m)$ diagram, with their fitted inner boundaries highlighted. Panel (c) compares the eccentricity distributions for the RV, \citet{Mayor_2011}, and CLS samples. For both the \citet{Mayor_2011} and CLS samples, the inverse detection completeness is applied as weights when evaluating the density distribution. A qualitative comparison of these three samples shows good agreement in the demographic features discussed in this paper.

Specifically, the slope and position of the inner boundaries exhibit quantitative agreement within an acceptable range. The $k$ values (slopes) for the inner boundaries are 2.99, 2.9, and 3.37 for the RV, \citet{Mayor_2011}, and CLS samples, respectively. The semi-major axis values of the largest mass bin (indicated by the upper-most red cross) are 0.69, 0.72, and 0.87 au for the RV, \citet{Mayor_2011}, and CLS samples. The density peaks in each mass bin (represented by the vertical yellow bars) are positioned further out, especially for larger masses, in the CLS sample compared to the other two samples. This discrepancy is likely due to the detection efficiency of planets on distant orbits ($a \gtrsim 10~$au) in the CLS survey. However, the features of the inner boundary remain consistent with those of the other two samples, as they are relatively insensitive to detection efficiency. Finally, Panel (c) shows that the eccentricity distributions are in good agreement across the three samples.
Thus, we conclude that our findings are insensitive to detection biases and do not depend on the choice of survey.

\section{Stellar mass and metallicity} \label{appendix:RV_star}

\begin{figure*}
    \centering
    \includegraphics[width=0.98\textwidth]{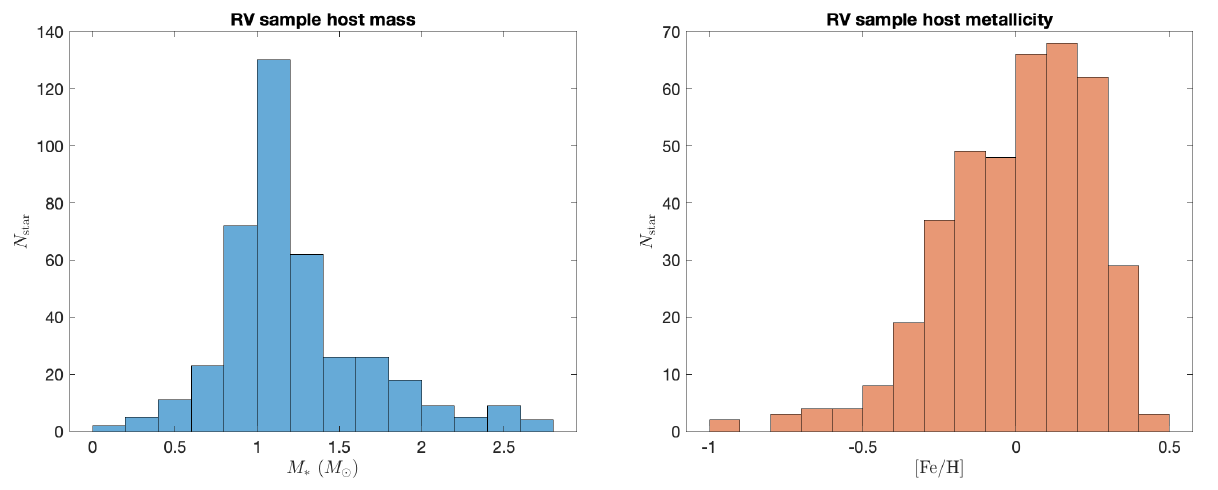}
    \caption{The mass and metallicity distribution of the 402 stars in the RV sample.}
    \label{fig:RV_mass_metallicity}
\end{figure*}

Figure \ref{fig:RV_mass_metallicity} shows the distribution of the mass and metallicity of the 402 stars in the RV sample, as well as in the stellar input of Models 1-6 and 11.

\bibliography{aastex631/references}{}

\begin{thebibliography}{}
\expandafter\ifx\csname natexlab\endcsname\relax\def\natexlab#1{#1}\fi
\providecommand{\url}[1]{\href{#1}{#1}}
\providecommand{\dodoi}[1]{doi:~\href{http://doi.org/#1}{\nolinkurl{#1}}}
\providecommand{\doeprint}[1]{\href{http://ascl.net/#1}{\nolinkurl{http://ascl.net/#1}}}
\providecommand{\doarXiv}[1]{\href{https://arxiv.org/abs/#1}{\nolinkurl{https://arxiv.org/abs/#1}}}

\bibitem[{{Artymowicz}(1993)}]{Artymowicz_1993}
{Artymowicz}, P. 1993, \apj, 419, 155, \dodoi{10.1086/173469}

\bibitem[{{Bai}(2017)}]{Bai_2017}
{Bai}, X.-N. 2017, \apj, 845, 75, \dodoi{10.3847/1538-4357/aa7dda}

\bibitem[{{Batygin} \& {Morbidelli}(2023)}]{Batygin_2023}
{Batygin}, K., \& {Morbidelli}, A. 2023, Nature Astronomy, 7, 330, \dodoi{10.1038/s41550-022-01850-5}

\bibitem[{{Bitsch} {et~al.}(2019){Bitsch}, {Izidoro}, {Johansen}, {Raymond}, {Morbidelli}, {Lambrechts}, \& {Jacobson}}]{Bitsch_2019}
{Bitsch}, B., {Izidoro}, A., {Johansen}, A., {et~al.} 2019, \aap, 623, A88, \dodoi{10.1051/0004-6361/201834489}

\bibitem[{{Bitsch} {et~al.}(2015){Bitsch}, {Lambrechts}, \& {Johansen}}]{Bitsch_2015}
{Bitsch}, B., {Lambrechts}, M., \& {Johansen}, A. 2015, \aap, 582, A112, \dodoi{10.1051/0004-6361/201526463}

\bibitem[{Chambers(2018)}]{Chambers_2018}
Chambers, J. 2018, The Astrophysical Journal, 865, 30, \dodoi{10.3847/1538-4357/aada09}

\bibitem[{{Charnoz} {et~al.}(2021){Charnoz}, {Avice}, {Hyodo}, {Pignatale}, \& {Chaussidon}}]{Charnoz_2021}
{Charnoz}, S., {Avice}, G., {Hyodo}, R., {Pignatale}, F.~C., \& {Chaussidon}, M. 2021, \aap, 652, A35, \dodoi{10.1051/0004-6361/202038797}

\bibitem[{{Chatterjee} \& {Tan}(2014)}]{Chatterjee_2014}
{Chatterjee}, S., \& {Tan}, J.~C. 2014, \apj, 780, 53, \dodoi{10.1088/0004-637X/780/1/53}

\bibitem[{{Crida} {et~al.}(2006){Crida}, {Morbidelli}, \& {Masset}}]{Crida_2006}
{Crida}, A., {Morbidelli}, A., \& {Masset}, F. 2006, \icarus, 181, 587, \dodoi{10.1016/j.icarus.2005.10.007}

\bibitem[{{Dr{\k{a}}{\.z}kowska} \& {Alibert}(2017)}]{Drazkowska_2017}
{Dr{\k{a}}{\.z}kowska}, J., \& {Alibert}, Y. 2017, \aap, 608, A92, \dodoi{10.1051/0004-6361/201731491}

\bibitem[{{Dr{\k{a}}{\.z}kowska} {et~al.}(2016){Dr{\k{a}}{\.z}kowska}, {Alibert}, \& {Moore}}]{Drazkowska_2016}
{Dr{\k{a}}{\.z}kowska}, J., {Alibert}, Y., \& {Moore}, B. 2016, \aap, 594, A105, \dodoi{10.1051/0004-6361/201628983}

\bibitem[{{Dr{\k{a}}{\.z}kowska} \& {Dullemond}(2018)}]{Drazkowska_2018}
{Dr{\k{a}}{\.z}kowska}, J., \& {Dullemond}, C.~P. 2018, \aap, 614, A62, \dodoi{10.1051/0004-6361/201732221}

\bibitem[{{Emsenhuber} {et~al.}(2021{\natexlab{a}}){Emsenhuber}, {Mordasini}, {Burn}, {Alibert}, {Benz}, \& {Asphaug}}]{Emsenhuber_2021_I}
{Emsenhuber}, A., {Mordasini}, C., {Burn}, R., {et~al.} 2021{\natexlab{a}}, \aap, 656, A69, \dodoi{10.1051/0004-6361/202038553}

\bibitem[{{Emsenhuber} {et~al.}(2021{\natexlab{b}}){Emsenhuber}, {Mordasini}, {Burn}, {Alibert}, {Benz}, \& {Asphaug}}]{Emsenhuber_2021_II}
---. 2021{\natexlab{b}}, \aap, 656, A70, \dodoi{10.1051/0004-6361/202038863}

\bibitem[{Feng {et~al.}(2022)Feng, Butler, Vogt, Clement, Tinney, Cui, Aizawa, Jones, Bailey, Burt, Carter, Crane, Dotti, Holden, Ma, Ogihara, Oppenheimer, OâToole, Shectman, Wittenmyer, Wang, Wright, \& Xuan}]{Feng_2022}
Feng, F., Butler, R.~P., Vogt, S.~S., {et~al.} 2022, The Astrophysical Journal Supplement Series, 262, 21, \dodoi{10.3847/1538-4365/ac7e57}

\bibitem[{Fernandes {et~al.}(2019)Fernandes, Mulders, Pascucci, Mordasini, \& Emsenhuber}]{Fernandes_2019}
Fernandes, R.~B., Mulders, G.~D., Pascucci, I., Mordasini, C., \& Emsenhuber, A. 2019, The Astrophysical Journal, 874, 81, \dodoi{10.3847/1538-4357/ab0300}

\bibitem[{Fulton {et~al.}(2021)Fulton, Rosenthal, Hirsch, Isaacson, Howard, Dedrick, Sherstyuk, Blunt, Petigura, Knutson, Behmard, Chontos, Crepp, Crossfield, Dalba, Fischer, Henry, Kane, Kosiarek, Marcy, Rubenzahl, Weiss, \& Wright}]{Fulton_2021}
Fulton, B.~J., Rosenthal, L.~J., Hirsch, L.~A., {et~al.} 2021, The Astrophysical Journal Supplement Series, 255, 14, \dodoi{10.3847/1538-4365/abfcc1}

\bibitem[{{Garaud} \& {Lin}(2007)}]{Garaud_2007}
{Garaud}, P., \& {Lin}, D.~N.~C. 2007, \apj, 654, 606, \dodoi{10.1086/509041}

\bibitem[{{Goldreich} \& {Tremaine}(1980)}]{Goldreich_1980}
{Goldreich}, P., \& {Tremaine}, S. 1980, \apj, 241, 425, \dodoi{10.1086/158356}

\bibitem[{{Guilera} {et~al.}(2020){Guilera}, {S{\'a}ndor}, {Ronco}, {Venturini}, \& {Miller Bertolami}}]{Guilera_2020}
{Guilera}, O.~M., {S{\'a}ndor}, Z., {Ronco}, M.~P., {Venturini}, J., \& {Miller Bertolami}, M.~M. 2020, \aap, 642, A140, \dodoi{10.1051/0004-6361/202038458}

\bibitem[{Hansen(2009)}]{Hansen_2009}
Hansen, B. M.~S. 2009, The Astrophysical Journal, 703, 1131, \dodoi{10.1088/0004-637X/703/1/1131}

\bibitem[{Hayashi(1981)}]{Hayashi_1981}
Hayashi, C. 1981, Progress of Theoretical Physics Supplement, 70, 35, \dodoi{10.1143/PTPS.70.35}

\bibitem[{{Hori} \& {Ikoma}(2010)}]{Hori_2010}
{Hori}, Y., \& {Ikoma}, M. 2010, \apj, 714, 1343, \dodoi{10.1088/0004-637X/714/2/1343}

\bibitem[{Hori \& Ogihara(2020)}]{Hori_2020}
Hori, Y., \& Ogihara, M. 2020, The Astrophysical Journal, 889, 77, \dodoi{10.3847/1538-4357/ab6168}

\bibitem[{{Hsieh} \& {Lin}(2020)}]{Hsieh_2020}
{Hsieh}, H.-F., \& {Lin}, M.-K. 2020, \mnras, 497, 2425, \dodoi{10.1093/mnras/staa2115}

\bibitem[{{Hyodo} {et~al.}(2019){Hyodo}, {Ida}, \& {Charnoz}}]{Hyodo_2019}
{Hyodo}, R., {Ida}, S., \& {Charnoz}, S. 2019, \aap, 629, A90, \dodoi{10.1051/0004-6361/201935935}

\bibitem[{{Ida} \& {Guillot}(2016)}]{Ida_2016}
{Ida}, S., \& {Guillot}, T. 2016, \aap, 596, L3, \dodoi{10.1051/0004-6361/201629680}

\bibitem[{Ida \& Lin(2004{\natexlab{a}})}]{Ida_2004a}
Ida, S., \& Lin, D. N.~C. 2004{\natexlab{a}}, The Astrophysical Journal, 604, 388, \dodoi{10.1086/381724}

\bibitem[{Ida \& Lin(2004{\natexlab{b}})}]{Ida_2004b}
---. 2004{\natexlab{b}}, The Astrophysical Journal, 616, 567, \dodoi{10.1086/424830}

\bibitem[{Ida \& Lin(2005)}]{Ida_2005}
---. 2005, The Astrophysical Journal, 626, 1045, \dodoi{10.1086/429953}

\bibitem[{{Ida} \& {Lin}(2008)}]{Ida_2008a}
{Ida}, S., \& {Lin}, D.~N.~C. 2008, \apj, 673, 487, \dodoi{10.1086/523754}

\bibitem[{{Ida} \& {Lin}(2010)}]{Ida_2010}
---. 2010, \apj, 719, 810, \dodoi{10.1088/0004-637X/719/1/810}

\bibitem[{{Ida} {et~al.}(2013){Ida}, {Lin}, \& {Nagasawa}}]{Ida_2013}
{Ida}, S., {Lin}, D.~N.~C., \& {Nagasawa}, M. 2013, \apj, 775, 42, \dodoi{10.1088/0004-637X/775/1/42}

\bibitem[{{Ida} {et~al.}(2018){Ida}, {Tanaka}, {Johansen}, {Kanagawa}, \& {Tanigawa}}]{Ida_2018}
{Ida}, S., {Tanaka}, H., {Johansen}, A., {Kanagawa}, K.~D., \& {Tanigawa}, T. 2018, \apj, 864, 77, \dodoi{10.3847/1538-4357/aad69c}

\bibitem[{{Ikoma} {et~al.}(2000){Ikoma}, {Nakazawa}, \& {Emori}}]{Ikoma_2000}
{Ikoma}, M., {Nakazawa}, K., \& {Emori}, H. 2000, \apj, 537, 1013, \dodoi{10.1086/309050}

\bibitem[{{Izidoro} {et~al.}(2022){Izidoro}, {Dasgupta}, {Raymond}, {Deienno}, {Bitsch}, \& {Isella}}]{Izidoro_2022}
{Izidoro}, A., {Dasgupta}, R., {Raymond}, S.~N., {et~al.} 2022, Nature Astronomy, 6, 357, \dodoi{10.1038/s41550-021-01557-z}

\bibitem[{{Izidoro} {et~al.}(2015){Izidoro}, {Raymond}, {Morbidelli}, {Hersant}, \& {Pierens}}]{Izidoro_2015}
{Izidoro}, A., {Raymond}, S.~N., {Morbidelli}, A., {Hersant}, F., \& {Pierens}, A. 2015, \apjl, 800, L22, \dodoi{10.1088/2041-8205/800/2/L22}

\bibitem[{{Johansen} {et~al.}(2019){Johansen}, {Ida}, \& {Brasser}}]{Johansen_2019}
{Johansen}, A., {Ida}, S., \& {Brasser}, R. 2019, \aap, 622, A202, \dodoi{10.1051/0004-6361/201834071}

\bibitem[{{Kanagawa} {et~al.}(2018){Kanagawa}, {Tanaka}, \& {Szuszkiewicz}}]{Kanagawa_2018}
{Kanagawa}, K.~D., {Tanaka}, H., \& {Szuszkiewicz}, E. 2018, \apj, 861, 140, \dodoi{10.3847/1538-4357/aac8d9}

\bibitem[{{Kane} \& {Wittenmyer}(2024)}]{Kane_2024}
{Kane}, S.~R., \& {Wittenmyer}, R.~A. 2024, \apjl, 962, L21, \dodoi{10.3847/2041-8213/ad2463}

\bibitem[{{Kokubo} \& {Ida}(2002)}]{Kokubo_Ida_2002}
{Kokubo}, E., \& {Ida}, S. 2002, \apj, 581, 666, \dodoi{10.1086/344105}

\bibitem[{{Lambrechts} \& {Johansen}(2012)}]{Lambrechts_2012}
{Lambrechts}, M., \& {Johansen}, A. 2012, \aap, 544, A32, \dodoi{10.1051/0004-6361/201219127}

\bibitem[{{Laughlin} {et~al.}(2004){Laughlin}, {Steinacker}, \& {Adams}}]{Laughlin_2004}
{Laughlin}, G., {Steinacker}, A., \& {Adams}, F.~C. 2004, \apj, 608, 489, \dodoi{10.1086/386316}

\bibitem[{{Li} {et~al.}(2005){Li}, {Li}, {Koller}, {Wendroff}, {Liska}, {Orban}, {Liang}, \& {Lin}}]{Li_2005}
{Li}, H., {Li}, S., {Koller}, J., {et~al.} 2005, \apj, 624, 1003, \dodoi{10.1086/429367}

\bibitem[{{Lichtenberg} {et~al.}(2021){Lichtenberg}, {Dr{\k{a}}{\.z}kowska}, {Sch{\"o}nb{\"a}chler}, {Golabek}, \& {Hands}}]{Lichtenberg_2021}
{Lichtenberg}, T., {Dr{\k{a}}{\.z}kowska}, J., {Sch{\"o}nb{\"a}chler}, M., {Golabek}, G.~J., \& {Hands}, T.~O. 2021, Science, 371, 365, \dodoi{10.1126/science.abb3091}

\bibitem[{{Lin} \& {Papaloizou}(1993)}]{Lin_1993}
{Lin}, D.~N.~C., \& {Papaloizou}, J.~C.~B. 1993, in Protostars and Planets III, ed. E.~H. {Levy} \& J.~I. {Lunine}, 749

\bibitem[{{Liu} {et~al.}(2019){Liu}, {Lambrechts}, {Johansen}, \& {Liu}}]{Liu_2019b}
{Liu}, B., {Lambrechts}, M., {Johansen}, A., \& {Liu}, F. 2019, \aap, 632, A7, \dodoi{10.1051/0004-6361/201936309}

\bibitem[{{Lynden-Bell} \& {Pringle}(1974)}]{Lyndel-Bell_1974}
{Lynden-Bell}, D., \& {Pringle}, J.~E. 1974, \mnras, 168, 603, \dodoi{10.1093/mnras/168.3.603}

\bibitem[{{Masset} {et~al.}(2006{\natexlab{a}}){Masset}, {D'Angelo}, \& {Kley}}]{Masset_2006a}
{Masset}, F.~S., {D'Angelo}, G., \& {Kley}, W. 2006{\natexlab{a}}, \apj, 652, 730, \dodoi{10.1086/507515}

\bibitem[{{Masset} {et~al.}(2006{\natexlab{b}}){Masset}, {Morbidelli}, {Crida}, \& {Ferreira}}]{Masset_2006b}
{Masset}, F.~S., {Morbidelli}, A., {Crida}, A., \& {Ferreira}, J. 2006{\natexlab{b}}, \apj, 642, 478, \dodoi{10.1086/500967}

\bibitem[{{Matsumura} {et~al.}(2021){Matsumura}, {Brasser}, \& {Ida}}]{Matsumura_2021}
{Matsumura}, S., {Brasser}, R., \& {Ida}, S. 2021, \aap, 650, A116, \dodoi{10.1051/0004-6361/202039210}

\bibitem[{Matsumura {et~al.}(2013)Matsumura, Ida, \& Nagasawa}]{Matsumura_2013}
Matsumura, S., Ida, S., \& Nagasawa, M. 2013, The Astrophysical Journal, 767, 129, \dodoi{10.1088/0004-637X/767/2/129}

\bibitem[{{Mayor} {et~al.}(2011){Mayor}, {Marmier}, {Lovis}, {Udry}, {S{\'e}gransan}, {Pepe}, {Benz}, {Bertaux}, {Bouchy}, {Dumusque}, {Lo Curto}, {Mordasini}, {Queloz}, \& {Santos}}]{Mayor_2011}
{Mayor}, M., {Marmier}, M., {Lovis}, C., {et~al.} 2011, arXiv e-prints, arXiv:1109.2497, \dodoi{10.48550/arXiv.1109.2497}

\bibitem[{{Morbidelli} {et~al.}(2022){Morbidelli}, {Bailli{\'e}}, {Batygin}, {Charnoz}, {Guillot}, {Rubie}, \& {Kleine}}]{Morbidelli_2022}
{Morbidelli}, A., {Bailli{\'e}}, K., {Batygin}, K., {et~al.} 2022, Nature Astronomy, 6, 72, \dodoi{10.1038/s41550-021-01517-7}

\bibitem[{{Morbidelli} \& {Nesvorny}(2012)}]{Morbidelli_2012}
{Morbidelli}, A., \& {Nesvorny}, D. 2012, \aap, 546, A18, \dodoi{10.1051/0004-6361/201219824}

\bibitem[{Nagasawa {et~al.}(2008)Nagasawa, Ida, \& Bessho}]{Nagasawa_2008}
Nagasawa, M., Ida, S., \& Bessho, T. 2008, The Astrophysical Journal, 678, 498, \dodoi{10.1086/529369}

\bibitem[{{NASA Exoplanet Archive}(2023)}]{PSCompPars}
{NASA Exoplanet Archive}. 2023, Planetary Systems Composite Parameters, Version: 2023-10-30 HH:MM,  NExScI-Caltech/IPAC, \dodoi{10.26133/NEA13}

\bibitem[{{Ndugu} {et~al.}(2018){Ndugu}, {Bitsch}, \& {Jurua}}]{Ndugu_2018}
{Ndugu}, N., {Bitsch}, B., \& {Jurua}, E. 2018, \mnras, 474, 886, \dodoi{10.1093/mnras/stx2815}

\bibitem[{{Nelson} \& {Papaloizou}(2004)}]{Nelson_2004}
{Nelson}, R.~P., \& {Papaloizou}, J. C.~B. 2004, \mnras, 350, 849, \dodoi{10.1111/j.1365-2966.2004.07406.x}

\bibitem[{{Ogihara} {et~al.}(2015{\natexlab{a}}){Ogihara}, {Kobayashi}, {Inutsuka}, \& {Suzuki}}]{Ogihara_2015a}
{Ogihara}, M., {Kobayashi}, H., {Inutsuka}, S.-i., \& {Suzuki}, T.~K. 2015{\natexlab{a}}, \aap, 579, A65, \dodoi{10.1051/0004-6361/201525636}

\bibitem[{{Ogihara} {et~al.}(2018){Ogihara}, {Kokubo}, {Suzuki}, \& {Morbidelli}}]{Ogihara_2018}
{Ogihara}, M., {Kokubo}, E., {Suzuki}, T.~K., \& {Morbidelli}, A. 2018, \aap, 612, L5, \dodoi{10.1051/0004-6361/201832654}

\bibitem[{Ogihara {et~al.}(2020)Ogihara, Kunitomo, \& Hori}]{Ogihara_2020}
Ogihara, M., Kunitomo, M., \& Hori, Y. 2020, The Astrophysical Journal, 899, 91, \dodoi{10.3847/1538-4357/aba75e}

\bibitem[{{Ogihara} {et~al.}(2015{\natexlab{b}}){Ogihara}, {Morbidelli}, \& {Guillot}}]{Ogihara_2015b}
{Ogihara}, M., {Morbidelli}, A., \& {Guillot}, T. 2015{\natexlab{b}}, \aap, 584, L1, \dodoi{10.1051/0004-6361/201527117}

\bibitem[{{Ogihara} {et~al.}(2024){Ogihara}, {Morbidelli}, \& {Kunitomo}}]{Ogihara_2024}
{Ogihara}, M., {Morbidelli}, A., \& {Kunitomo}, M. 2024, \apj, 972, 181, \dodoi{10.3847/1538-4357/ad65d5}

\bibitem[{Oka {et~al.}(2011)Oka, Nakamoto, \& Ida}]{Oka_2011}
Oka, A., Nakamoto, T., \& Ida, S. 2011, The Astrophysical Journal, 738, 141, \dodoi{10.1088/0004-637X/738/2/141}

\bibitem[{{Ormel} \& {Klahr}(2010)}]{Ormel_2010}
{Ormel}, C.~W., \& {Klahr}, H.~H. 2010, \aap, 520, A43, \dodoi{10.1051/0004-6361/201014903}

\bibitem[{{Paardekooper} {et~al.}(2011){Paardekooper}, {Baruteau}, \& {Kley}}]{Paardekooper_2011}
{Paardekooper}, S.~J., {Baruteau}, C., \& {Kley}, W. 2011, \mnras, 410, 293, \dodoi{10.1111/j.1365-2966.2010.17442.x}

\bibitem[{{Rein} \& {Liu}(2012)}]{Rein_2012}
{Rein}, H., \& {Liu}, S.~F. 2012, \aap, 537, A128, \dodoi{10.1051/0004-6361/201118085}

\bibitem[{{Rein} \& {Spiegel}(2015)}]{Rein_2015_IAS15}
{Rein}, H., \& {Spiegel}, D.~S. 2015, \mnras, 446, 1424, \dodoi{10.1093/mnras/stu2164}

\bibitem[{{Ros} \& {Johansen}(2013)}]{Ros_2013}
{Ros}, K., \& {Johansen}, A. 2013, \aap, 552, A137, \dodoi{10.1051/0004-6361/201220536}

\bibitem[{Rosenthal {et~al.}(2023)Rosenthal, Howard, Knutson, \& Fulton}]{Rosenthal_2024}
Rosenthal, L.~J., Howard, A.~W., Knutson, H.~A., \& Fulton, B.~J. 2023, The Astrophysical Journal Supplement Series, 270, 1, \dodoi{10.3847/1538-4365/acffc0}

\bibitem[{Rosenthal {et~al.}(2022)Rosenthal, Knutson, Chachan, Dai, Howard, Fulton, Chontos, Crepp, Dalba, Henry, Kane, Petigura, Weiss, \& Wright}]{Rosenthal_2022}
Rosenthal, L.~J., Knutson, H.~A., Chachan, Y., {et~al.} 2022, The Astrophysical Journal Supplement Series, 262, 1, \dodoi{10.3847/1538-4365/ac7230}

\bibitem[{{Schoonenberg} \& {Ormel}(2017)}]{Schoonenberg_2017}
{Schoonenberg}, D., \& {Ormel}, C.~W. 2017, \aap, 602, A21, \dodoi{10.1051/0004-6361/201630013}

\bibitem[{{Suzuki} {et~al.}(2010){Suzuki}, {Muto}, \& {Inutsuka}}]{Suzuki_2010}
{Suzuki}, T.~K., {Muto}, T., \& {Inutsuka}, S.-i. 2010, \apj, 718, 1289, \dodoi{10.1088/0004-637X/718/2/1289}

\bibitem[{{Tanaka} {et~al.}(2020){Tanaka}, {Murase}, \& {Tanigawa}}]{Tanaka_2020}
{Tanaka}, H., {Murase}, K., \& {Tanigawa}, T. 2020, \apj, 891, 143, \dodoi{10.3847/1538-4357/ab77af}

\bibitem[{{Tanaka} {et~al.}(2002){Tanaka}, {Takeuchi}, \& {Ward}}]{Tanaka_2002}
{Tanaka}, H., {Takeuchi}, T., \& {Ward}, W.~R. 2002, \apj, 565, 1257, \dodoi{10.1086/324713}

\bibitem[{{Ueda} {et~al.}(2021){Ueda}, {Ogihara}, {Kokubo}, \& {Okuzumi}}]{Ueda_2021}
{Ueda}, T., {Ogihara}, M., {Kokubo}, E., \& {Okuzumi}, S. 2021, \apjl, 921, L5, \dodoi{10.3847/2041-8213/ac2f3b}

\bibitem[{Woo {et~al.}(2023)Woo, Morbidelli, Grimm, Stadel, \& Brasser}]{Woo_2023}
Woo, J., Morbidelli, A., Grimm, S., Stadel, J., \& Brasser, R. 2023, Icarus, 396, 115497, \dodoi{https://doi.org/10.1016/j.icarus.2023.115497}

\bibitem[{Woo {et~al.}(2024)Woo, Nesvorný, Scora, \& Morbidelli}]{Woo_2024}
Woo, J., Nesvorný, D., Scora, J., \& Morbidelli, A. 2024, Icarus, 417, 116109, \dodoi{https://doi.org/10.1016/j.icarus.2024.116109}

\bibitem[{Yang {et~al.}(2023)Yang, Wu, Zheng, Ogihara, Guo, Ouyang, \& He}]{Yang_2023}
Yang, S., Wu, L., Zheng, Z., {et~al.} 2023, Icarus, 406, 115757, \dodoi{https://doi.org/10.1016/j.icarus.2023.115757}

\bibitem[{Zhou {et~al.}(2007)Zhou, Lin, \& Sun}]{Zhou_2007}
Zhou, J.-L., Lin, D. N.~C., \& Sun, Y.-S. 2007, The Astrophysical Journal, 666, 423, \dodoi{10.1086/519918}

\bibitem[{Zhu \& Wu(2018)}]{Zhu_2018}
Zhu, W., \& Wu, Y. 2018, The Astronomical Journal, 156, 92, \dodoi{10.3847/1538-3881/aad22a}

\end{thebibliography}
\bibliographystyle{aasjournal}


\end{document}